\documentclass[ preprint, showkeys, aps]{revtex4-1}
\usepackage{amsmath}
\usepackage{graphicx}
\usepackage{dcolumn}
\usepackage{bm}
\usepackage{float}
\usepackage{subfigure}
\usepackage{adjustbox}
\usepackage{xcolor}
\usepackage{tcolorbox}
\usepackage{siunitx}
\DeclareMathOperator{\erfc}{erfc}

\begin{document}

\title{Microtubule organization and cell geometry}
\author{Panayiotis Foteinopoulos}
\author{Bela M. Mulder}
\affiliation{Institute AMOLF, Science Park 104, 1098XG Amsterdam, the Netherlands}
\email{mulder@amolf.nl}
\date{\today }

\begin{abstract}
We present a systematic study of the influence of cell geometry on the orientational distribution of microtubules (MTs) nucleated from a single microtubule organizing center (MTOC). For simplicity we consider an elliptical cell geometry, a setting appropriate to a generic non-spherical animal cell. Within this context we introduce four models of increasing complexity, in each case introducing additional mechanisms that govern the interaction of the MTs with the cell boundary. In order, we consider the cases: MTs that can bind to the boundary with a fixed mean residence time (M0),  force-producing MTs that can slide on the boundary towards the cell poles (MS), MTs that interact with a generic polarity factor that is transported and deposited at the boundary, and which in turn stabilizes the MTs at the boundary (MP), and a final model in which both sliding and stabilization by polarity factors is taken into account (MSP). In the baseline model (M0), the exponential length distribution of MTs causes most of the interactions at the cell boundary to occur along the shorter transverse direction in the cell, leading to transverse biaxial order. MT sliding (MS) is able to reorient the main axis of this biaxial order along the longitudinal axis. The polarization mechanism introduced in MP and MSP overrules the geometric bias towards bipolar order observed in M0 and MS, and allows the establishment of unipolar order either along the short- (MP) or the long cell axis (MSP). The behavior of the latter two models can be qualitatively reproduced by a very simple toy model with discrete MT orientations. 
\end{abstract}

\maketitle

\affiliation{Living Matter Department, Institute AMOLF, Science Park 104,
1098XG Amsterdam, the Netherlands}




\section{\label{sec:intro}Introduction}
In a typical animal cell, microtubules play a major role
in the intracellular bi-directional trafficking of molecules from the
vicinity of the cell nucleus to the cell periphery (for a recent review see \cite{Barlan2017Microtubule-BasedOrganelles}). In support of
this function, the microtubule cytoskeleton generically displays a radial spatial organisation governed
by a single microtubule organizing center (MTOC) which localizes to the
nuclear envelope. The MTOC is the locus of protein complexes, such as $\gamma$%
-tubulin ring complexes, which serve to nucleate new MTs (for a recent
review see \cite{Wu2017Microtubule-organizingCenters}). These MTs
subsequently grow outward towards the cell periphery in roughly
uniformly distributed directions. MTOCs, depending on cell type, can
support up to hundreds of MTs \cite{Oriola2015TheSpindle}.

The question what happens when the MTs still connected to the MTOC reach the
cell boundary has to date mostly been considered in the context of two
biologically relevant phenomena. The first phenomenon is the role of these \emph{bound}
MTs play in the localization of the MTOC itself. In the seminal work by Tran et al.\ \cite%
{Tran2001}, it was shown that MTs, by virtue of the polymerization forces
they exert when in contact with the cell membrane, are able to robustly
center the nucleus in fission yeast to which they are connected the
so-called by spindle pole bodies on the nuclear envelope that act as multiple MTOCs in this case.
Later work focused on the question of the positioning of in-vitro
reconstituted MT-asters, the star-like structure of MTs emerging from an MTOC,  in lithographically produced microchambers \cite%
{Laan2012,Laan2012a}, with a view of understanding spindle
pole positioning in cells (see e.g.\ \cite{Meaders2020AType,Jimenez2021Acto-myosinPosition}). An important insight
gained from this work, is that in more spherical cells pushing forces do not
provide a robust centering. Centering is only obtained if pulling forces,
exerted by e.g.\ cortical dynein motor proteins, are also at play. Related is the
observation that in a non-spherical cell the polymerization force exerted by a growing MT at the membrane inevitably has a tangential component causing MTs to slide, and thus reshaping the MT distribution within the cell \cite{pavin2012,ma2014}. 

The second phenomenon is the possible role of MTs in setting up and maintaining
cell polarity (for a recent review see e.g.\ \cite{Thompson2013}). The
paradigmatic example of a system of this type was first discovered in
fission yeast, in which the polarity marker Tea1p is transported by the
motor protein Tea2p along longitudinally oriented MTs and subsequently
delivered on the apical membrane aided by the MT-end binding protein Mal3p 
\cite{Mata1997Tea1Cell,Brunner2000CLIP170-likeYeast}. As Tea1p is an example
of a so-called microtubule associated protein (MAP), a large class of
proteins that bind to MTs and are e.g.\ able to alter their dynamics,
it has been speculated that this could form the basis of a robust and
self-sustaining polarization mechanism in which MTs serve to transport
factors to the membrane, that subsequently stabilize them and hence provide
a positive feedback-loop maintaining their localization \cite%
{Recouvreux2016,Foteinopoulos2017APolarity}.

The previous work described above strongly focused on specific questions and
geometries, and it is fair to say that we do not yet have a comprehensive
understanding of how cell geometry influences the global distribution of MTs
given various types of potential interactions of the MTs with the cell
boundary. Here we aim to fill this gap by presenting a systematic study,
allowing both the geometry to vary, both in size relative to the mean length
of the MTs and in shape, as well as considering four distinct scenarios of
MT-boundary interactions of increasing complexity. Specifically, we consider
the following models: (M0) MTs that can bind to the boundary with a fixed
mean residence time, (MS) force-producing MTs that can slide on the boundary
towards the cell poles, (MP) MTs that interact with a generic polarity
factor that is transported and deposited at the boundary, and which in
turn stabilizes the MTs at the boundary, and (MSP) a final model in which
both sliding and stabilization by polarity factors is taken into account. In
all cases we study these models through stochastic simulations. However, for
model M0 we also obtain a full analytical solution, while for the two models
involving polarity (MP and MSP), we construct stylized toy-models that allow
us to rationalize the observed behaviour. For the models involving
force-induce MT sliding, we employ a recently developed force-production
mechanism that explicitly accounts for the effect of force on the speed of
growth and the residence time of MTs at the membrane \cite{Teapal2021ForcedCells}. The two main questions we focus on are (i) whether the MTs are predominantly organised along the longer longitudinal axis or the shorter transverse axis of the cell and (ii) whether the symmetries of the cell shape are imposed on the MT organisation, or whether this symmetry can be broken, yielding a polarized state.

To make this analysis tractable, we make a number of approximations. The
first is that we will only consider a single MTOC whose position is fixed to
the geometrical center of the cellular geometry. This implies that we assume
that an effective central positioning mechanism, such as discussed above,  is already in place, which is \emph{not} perturbed
by the additional mechanisms of MT-boundary interactions we implement. This
assumption allow us to more straightforwardly disentangle the
differential effects on the global organisation of the MTs due to
differences in geometry and/or boundary conditions. Secondly, we restrict
ourselves to a 2D elliptical cell geometry. The choice for 2D affords
computational advantages that allow for a much more extensive range of
conditions to be addressed through stochastic simulations than would be
feasible in 3D. Moreover, experience with other systems has shown that one
can expect results that are good guide for the behaviour in the
corresponding 3D geometry with cylindrical symmetry obtained by rotating the 2D shape
along the longitudinal axis. In that light, the choice for an elliptical
shape is one of convenience, both analytically and computationally, and a
generic proxy for a smooth non-spherical cell geometry.

The paper is organised as follows: in Section \ref{sec:framework} we introduce our modelling framework, the cell geometry, the observables used (\ref{sec:geom_dynamics}), and introduce the four models we consider (\ref{sec:models}). In Section \ref{sec:results} we collect our results on the four models in order: M0 (\ref{sec:M0}), MS (\ref{sec:MS}), MP (\ref{sec:MP}) and MSP (\ref{sec:MSP}). In Section \ref{sec:toymodel} we rationalize and compare the results on models MP and MSP with the aid of a simple toy model. We end with a discussion in Section \ref{sec:discussion}. In three appendices we collect a number of technical details on the derivation of the MT dynamical equations underpinning model M0 (\ref{app:model_M0}), the implementation of the polarization mechanism in the presence of force production in model MSP (\ref{app:r+_MSP}), and a in-depth description of our toy model (\ref{app:toy}).

\section{Modelling framework}\label{sec:framework}

\subsection{Cell geometry and MT dynamics}

\label{sec:geom_dynamics}

We consider a point-like idealized MTOC located at the center of an elliptical
cell, with a major axis of half-length $b$, and minor axis of half-length $%
a<b$. Orientation of MTs in this geometry are specified by the azimuthal
angle $\varphi $, defined with respect to the positive major half-axis. We will call the direction along the major axis \emph{longitudinal} and the one along the minor axis \emph{transverse}.

Each MT is nucleated by a nucleation complex located at the MTOC, and we fix
the total number of these complexes, and hence the total number of MTs in
the system, to be $M$. If a nucleation site is unoccupied it will nucleate a
new growing MT with rate $r_{\mathrm{n}}$. It is conceptually convenient to consider
an unoccupied nucleation site as a \emph{dormant} MT, waiting to be
nucleated. We will consider two assumptions on the distribution of
nucleation angles, both consistent with isotropicity of the overall
nucleation pattern. The first, the \emph{homogeneous} scenario, assumes each
nucleation complex to fixedly point in a given direction, and that these
directions have constant angular density $m$, and hence $M=2\pi m$. The
second, the \emph{random} scenario, does not assume a fixed orientation for
the nucleation complexes, but has them fire in a randomly selected direction.

Once a MT has been nucleated, it follows the standard MT dynamical instability
model \cite{Dogterom1993}, with growth speed $v_{+},$ shrinking speed $v_{-}$%
, catastrophe rate $r_{+}$ and rescue rate $r_{-}$. When a MT hits the cell
boundary it stalls, remaining there until it detaches by switching to the
shrinking state with a rate $r_{\mathrm{u}}$. The length of a MT when it hits the
boundary in a given direction is given by $l_{\mathrm{b}}(\varphi )$, the latter
function encoding all the relevant information about the shape of the cell.
For the ellipse this length is given by 
\begin{equation}
l_{\mathrm{b}}(\varphi )=\frac{ab}{\sqrt{a^{2}\cos ^{2}\varphi +b^{2}\sin ^{2}\varphi 
}}
\end{equation}%
We focus on the steady state of these systems, in which the orientational
distribution of MTs can be described by the following quantities:

\begin{itemize}
\item $m_{0}(\varphi )$: The density (per angle) of \emph{dormant} MTs
pointing in the direction $\varphi $ at time $t$ in the homogeneous
nucleation scenario.

\item $M_{0}$: The number of dormant MTs in the random nucleation scenario.

\item $m_{+}(l,\varphi )$: The density (per angle per unit length) of \emph{%
growing} MTs of length $l$.

\item $m_{-}(l,\varphi )$: The density (per angle per unit length) of \emph{%
shrinking} MTs of length $l$.

\item $m_{\mathrm{a}}(l,\varphi )=m_{+}(l,\varphi )+m_{-}(l,\varphi )$: The density
(per angle per unit length) of \emph{active} MTs of length $l$

\item $m_{\mathrm{b}}(\varphi )$: The density (per angle) of MTs \emph{bound} to the
surface.
\end{itemize}

We also consider the associated length densities 
\begin{align}
L_{\mathrm{a}}(\varphi )& =\int_{0}^{\infty }\mathrm{d}l\,l\,m_{\mathrm{a}}(l,\varphi ) \\
L_{\mathrm{b}}(\varphi )& =l_{\mathrm{b}}(\varphi )m_{\mathrm{b}}(\varphi ) \\
L(\varphi )& =L_{\mathrm{a}}(\varphi )+L_{\mathrm{b}}(\varphi )
\end{align}%
To characterize the degree of orientational ordering of the MTs, we use two
order parameters. The first measures the degree of \emph{polar} ordering in
the frame of the cell geometry. It is defined as 
\begin{equation}
\mathbf{S}_{1}=(\langle \cos \varphi \rangle ,\langle \sin \varphi \rangle )
\label{eq:S1_eq}
\end{equation}%
where throughout the equilibrium average is defined through 
\begin{equation}
\left\langle f(\varphi )\right\rangle =\frac{\int_{0}^{2\pi }{\mathrm{d}\varphi {\,}%
f(\varphi ){\,}l(\varphi )}}{\int_{0}^{2\pi }{\mathrm{d}\varphi {\,}l(\varphi )}}
\end{equation}%
i.e.\ we focus on the distribution of MT length, or equivalently tubulin
mass. When convenient, the scalar order parameter $S_{1}=|\mathbf{S}_{1}|$
can be used as a measure of the magnitude of polar ordering, irrespective of
its orientation. The second order parameter measures the degree of \emph{%
bipolar} ordering, and is defined as%
\begin{equation}
\mathbf{S}_{2}=\left( 
\begin{array}{cc}
\left\langle \cos 2\varphi \right\rangle & \left\langle \sin 2\varphi
\right\rangle \\ 
\left\langle \sin 2\varphi \right\rangle & -\left\langle \cos 2\varphi
\right\rangle%
\end{array}%
\right)
\end{equation}%
Here it is convenient to introduce the scalar order parameter $S_{2}=\left( 
\mathbf{S}_{2}\right) _{xx}=\left\langle \cos 2\varphi \right\rangle .$ When 
$S_{2}>0$ the ordering is predominantly along the major axis (\emph{%
longitudinal}), while for $S_{2}<0$ the ordering is along the minor axis (%
\emph{transverse}).

\subsection{Models}\label{sec:models}

\begin{figure}[th]
\centerline{\includegraphics[width=0.75%
\textwidth]{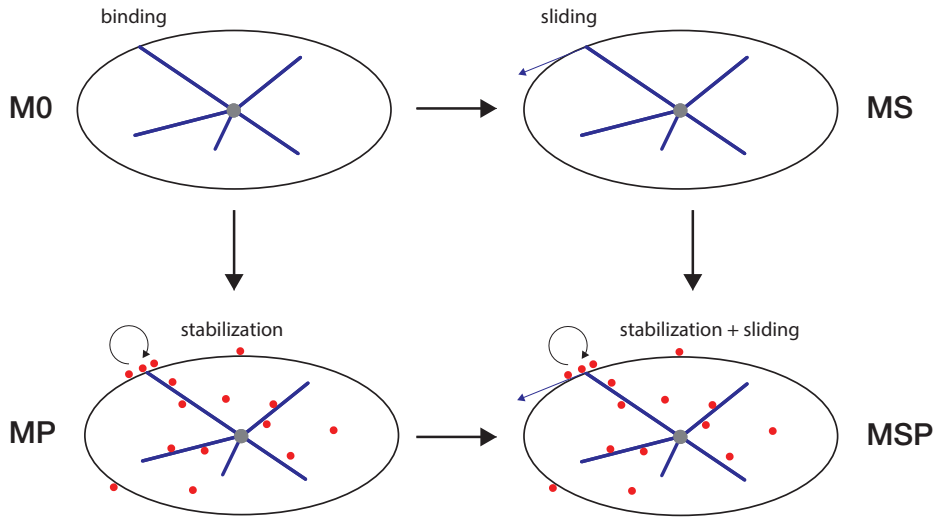}}
\caption{Schematic of the four models considered.}
\label{fig:4models}
\end{figure}

We will consider four models of increasing complexity describing the
interactions of the MTs with the boundary of our model cell. These models
are schematically illustrated in Fig.~\ref{fig:4models}.

\paragraph*{Model M0}

In this model, a growing MT that hits the boundary stalls there for a time
set by an \emph{unbinding rate}. When it unbinds it is in the shrinking
state.

\paragraph*{Model MS}

In this model, a growing MT that hits the boundary starts exerting a force.
The tangential component of the polymerization force exerted by the MT on the boundary then
causes the MT to \emph{slide} towards the nearest cell pole, and effect
counteracted by an (effective) friction force. At the same time, the rate at
which the MT grows is slowed and its catastrophe rate is increased, both in
a force dependent manner.

\paragraph*{Model MP}

In this model we introduce a species of effector molecules we dub \emph{%
polarity factors} (PFs). The PFs can bind to the MTs, which transport
them towards the cell boundary. Once deposited there, they diffuse away
and can reenter the cell interior at a given rate. The PFs in the boundary
stabilize the bound MTs in a density-dependent manner, in this way
creating a positive polarisation-inducing feedback loop.

\paragraph*{Model MSP}

In this final model both the force-induced sliding mechanism, and the
PF-induced polarization mechanism are active, yielding a model with maximal
coupling to the cell boundary and its geometry. 

\section{Models and results}\label{sec:results}

\subsection{Model M0: Fixed mean residence time at boundary}

\label{sec:M0} 
\begin{figure}[h]
\centerline{\includegraphics[width=0.75\textwidth]{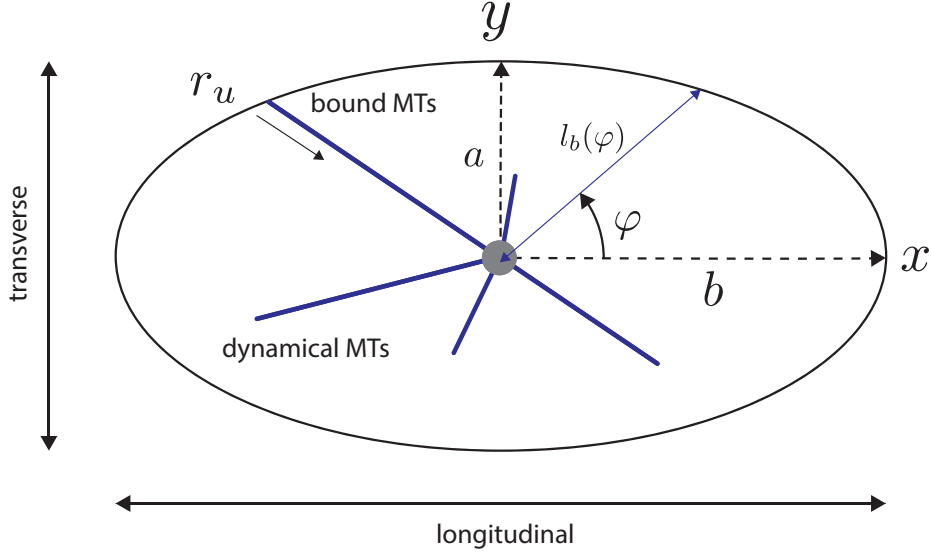}}
\caption{\textbf{Model M0}: Dynamical MTs are isotropically nucleated from a
fixed MTOC at the center of the cell. MTs remain bound to the membrane with
residence time set by the unbinding rate $r_{\mathrm{u}}$. Also indicated are the coordinate frame employed throughout and the two named directions characterizing the cell geometry. }
\label{fig:model_M0}
\end{figure}

In this first model, which will serve as the default case to
which the other models can be compared, the only effect on MTs reaching the
cell boundary is that they enter a \emph{bound} state, in which they are stalled.
Release from this bound state occurs at a constant unbinding rate $r_{\mathrm{u}}$, which does not depend on location. This mimics a generic nonspecific
interaction between the MT tip and the membrane, which by varying the value
of $r_{\mathrm{u}}$ ranges from repulsive  ($r_{\mathrm{u}} \gg 1$) to
(hyper)stabilizing ($r_{\mathrm{u}} \sim 0$). The model is schematically
illustrated in Fig.\ \ref{fig:model_M0}

\subsubsection{Analytical approach}

We focus on the behaviour of the relevant MT densities as introduced in Section \ref{sec:geom_dynamics} in the steady-state. The relevant equations follow from the time-dependent ones, which are presented in detail in Appendix \ref%
{app:model_M0}. For the growing, shrinking and bound MTs, respectively, we
have the balance equations 
\begin{align}
v_{+}\partial _{l}m_{+}(l,\varphi )& =r_{-}m_{-}(l,\varphi
)-r_{+}m_{+}(l,\varphi )  \label{eq:plus1} \\
-v_{-}\partial _{l}m_{-}(l,\varphi )& =-r_{-}m_{-}(l,\varphi
)+r_{+}m_{+}(l,\varphi )  \label{eq:minus1} \\
v_{+}m_{+}(l_{\mathrm{b}}(\varphi ),\varphi )& =r_{\mathrm{u}}m_{\mathrm{b}}(\varphi).
\label{eq:bound1}
\end{align}%
The behavior of the dormant MTs depends on the nucleation scenario, and we
have 
\begin{subequations}
\label{eq:dorm}
\begin{align}
r_{\mathrm{n}}m_{0}(\varphi )& =v_{-}m_{-}(l=0,\varphi ) \\
r_{\mathrm{n}}M_{0}& =\int_{0}^{2\pi }\mathrm{d}\varphi \,v_{-}m_{-}(l=0,\varphi ),
\end{align}%
where throughout the $a$-sublabelled equations will refer to the homogeneous
nucleation case, and the $b$-sublabelled ones to the random nucleation case.
These equations are supplemented by boundary conditions. At the cell
boundary we must have 
\end{subequations}
\begin{equation}
v_{-}m_{-}(l_{\mathrm{b}}(\varphi ))=r_{\mathrm{u}}m_{\mathrm{b}}(\varphi ),  \label{eq:BCb}
\end{equation}%
while at $l=0$ we have 
\begin{subequations}
\begin{align}
v_{+}(l=0,\varphi )& =r_{\mathrm{n}}m_{0}(\varphi )  \label{eq:BC0a} \\
v_{+}(l=0,\varphi )& =r_{\mathrm{n}}\frac{M_{0}}{2\pi }. \label{eq:BC0b}
\end{align}%
Adding Eqs.\ (\ref{eq:plus1}) and (\ref{eq:minus1}) gives 
\end{subequations}
\begin{equation}
\partial _{l}\{v_{+}m_{+}(l,\varphi )-v_{-}m_{-}(l,\varphi )\}=0
\label{eq:detbal}
\end{equation}%
Combining Eqs.\ (\ref{eq:bound1}) and (\ref{eq:BCb}), yields 
\begin{equation}
v_{+}m_{+}(l_{\mathrm{b}}(\varphi ),\varphi )=v_{-}m_{-}(l_{\mathrm{b}}(\varphi ),\varphi ),
\label{eq:balance}
\end{equation}%
which shows that the constant of integration in Eq.\ (\ref{eq:detbal})
vanishes and so we get 
\begin{equation}
v_{+}m_{+}(l,\varphi )=v_{-}m_{-}(l,\varphi ).
\end{equation}%
This allows us to eliminate $m_{-}(l,\varphi )$ in Eq.\ (\ref{eq:plus1}) and
solve it using either Eqs.\ (\ref{eq:BC0a}) or (\ref{eq:BC0b}), yielding 
\begin{subequations}
\label{eq:m_plus1}
\begin{align}
m_{+}(l,\varphi )& =\frac{r_{\mathrm{n}}}{v_{+}}m_{0}(\varphi )e^{-l/\bar{l}}
\label{eq:m_plus1hom} \\
m_{+}(l,\varphi )& =\frac{r_{\mathrm{n}}}{v_{+}}\frac{M_{0}}{2\pi }e^{-l/\bar{l}},
\label{eq:m_plus1ran}
\end{align}%
\end{subequations}
where 
\begin{equation}
\bar{l}=\left( \frac{r_{+}}{v_{+}}-\frac{r_{-}}{v_{-}}\right) ^{-1}
\label{eq:l_bar}
\end{equation}%
is the mean length of free MTs i.e.\ in the absence boundaries \cite{Dogterom1993}. From Eq.\ (%
\ref{eq:bound1}) we immediately get 
\begin{equation}
m_{\mathrm{b}}(\varphi )=\frac{v_{+}}{r_{\mathrm{u}}}m_{+}(l_{\mathrm{b}}(\varphi ),\varphi ).
\label{eq:Mb}
\end{equation}%
The final unknowns, pertaining to the dormant MTs, can now be obtained from
the appropriate conservation laws, which read 
\begin{subequations}
\begin{align}
m& =m_{0}(\varphi )+\int_{0}^{l_{\mathrm{b}}(\varphi )}{\mathrm{d}l{\,}\{m_{+}(l,\varphi
)+m_{-}(l,\varphi )\}}+m_{\mathrm{b}}(\varphi )  \label{eq:cons_hom} \\
M& =M_{0}+M_{\mathrm{a}}+M_{\mathrm{b}}, \label{eq:cons_ran}
\end{align}%
where 
\end{subequations}
\begin{align}
M_{\mathrm{a}}& =\int_{0}^{2\pi }\mathrm{d}\varphi \,\int_{0}^{L(\varphi )}{\mathrm{d}l{\,}%
\{m_{+}(l,\varphi )+m_{-}(l,\varphi )\}} \\
M_{\mathrm{b}}& =\int_{0}^{2\pi }\mathrm{d}\varphi \,m_{\mathrm{b}}(\varphi )
\end{align}%
are the total number of active and bound MTs in the system, respectively.
Inserting the results for $m_{+}(l,\varphi )$, $m_{-}(l,\varphi )$ and $%
m_{\mathrm{b}}(\varphi )$ in Eqs. \ (\ref{eq:cons_hom}) and (\ref{eq:cons_ran}) and introducing the convenient single MT ``partition function''
\begin{equation}
Z(\varphi )=1+r_{\mathrm{n}}\left( \frac{1}{v_{+}}+\frac{1}{v_{-}}\right) \bar{l}%
(1-e^{-l_{\mathrm{b}}(\varphi )/\bar{l}})+\frac{r_{\mathrm{n}}}{r_{\mathrm{u}}}e^{-l_{\mathrm{b}}(\varphi )/\bar{%
l}},
\end{equation}%
we find 
\begin{subequations}
\begin{align}
m& =m_{0}(\varphi )Z(\varphi ) \\
M& =M_{0}\,\tilde{Z},
\end{align}%
where throughout we use the tilde to denote the \emph{unweighted} average over
angles, i.e. 
\begin{equation*}
\tilde{f}\equiv \frac{1}{2\pi }\int_{0}^{2\pi }\mathrm{d}\varphi \,f(\varphi )
\end{equation*}
\end{subequations}
These results have a natural interpretation in terms of the following
timescales: $t_{0} =\frac{1}{r_{\mathrm{n}}}$, the mean residence time in the dormant
state, $\bar{t} =\left( \frac{1}{v_{+}}+\frac{1}{v_{-}}\right) \bar{l}$, the mean lifetime of an unperturbed MT, $t_{\mathrm{b}} =\frac{1}{r_{\mathrm{u}}}$, the mean residence time at the cell boundary, and the quantity 
\begin{equation}
F_{\mathrm{b}}(\varphi )=e^{-l_{\mathrm{b}}(\varphi )/\bar{l}},
\end{equation}%
which can interpreted as the probability that a MT reaches the boundary in
the direction $\varphi$. With these definitions we can write 
\begin{subequations}
\begin{align}
Z(\varphi ) & =\frac{t_{0}+\left( 1-F_{\mathrm{b}}(\varphi )\right) \bar{t}%
+F_{\mathrm{b}}(\varphi )t_{\mathrm{b}}}{t_{0}}\equiv \frac{t_{\mathrm{tot}}(\varphi )}{t_{0}} \\
\tilde{Z} &\equiv \frac{\tilde{t}_{\mathrm{tot}}}{t_{0}},
\end{align}%
\end{subequations}
where $t_{\mathrm{tot}}$ is interpreted as the total time spent in a single lifespan
of a MT starting in the dormant state ($t_{0}$), spending a time $t_{\mathrm{b}}$
bound to the surface with probability $F_{\mathrm{b}}$ and behaving as an unperturbed MT with probability $1-F_{\mathrm{b}}$. We now readily find the distribution of (i)
the dormant MTs 
\begin{subequations}
\begin{align}
m_{0}(\varphi )& =m\frac{t_{0}}{t_{\mathrm{tot}}(\varphi )} \\
M_{0}& =M\frac{t_{0}}{\tilde{t}_{\mathrm{tot}}},
\end{align}%
(ii) the active MTs 
\end{subequations}
\begin{subequations}
\begin{align}
m_{\mathrm{a}}(l,\varphi )& =m\frac{\bar{t}}{t_{\mathrm{tot}}(\varphi )}\frac{e^{-l/\bar{l}}}{%
\bar{l}} \\
m_{\mathrm{a}}(l,\varphi )& =\frac{m}{2\pi }\frac{\bar{t}}{\tilde{t}_{\mathrm{tot}}}\frac{%
e^{-l/\bar{l}}}{\bar{l}},
\end{align}%
and finally (iii) the bound MTs 
\end{subequations}
\begin{subequations}
\begin{align}
m_{\mathrm{b}}(\varphi )& =m\frac{t_{\mathrm{b}}}{t_{\mathrm{tot}}(\varphi )}F_{\mathrm{b}}(\varphi ) \\
m_{\mathrm{b}}(\varphi )& =\frac{m}{2\pi }\frac{t_{\mathrm{b}}}{\tilde{t}_{\mathrm{tot}}}F_{\mathrm{b}}(\varphi ).
\end{align}

Intriguingly, when $\bar{t}=t_{\mathrm{b}}$, we have $t_{\mathrm{tot}}(\varphi )=\tilde{t}%
_{\mathrm{tot}}=t_{0}+\bar{t}$. In this case the two nucleation scenarios lead to
exactly the same results. Intuitively this can be understood as follows. Due to the fact that the MT dynamics is Markovian,
The lifetime $\bar{t}$ is  also equal to the \emph{return time} i.e.\ the time it takes on average for
a growing MT to return to its initial length in the shrinking state. So,
whether a MT is kept at the boundary in the stalled state for a time $\bar{t}
$, or is free to propagate beyond the boundary and returning after a time 
$\bar{t}$, has no impact on the distribution within the boundary. From the
perspective of the MTOC the dynamics of departing and returning MTs is as if
they are launched into unbounded space in which no direction is favored,
which removes the distinction between the two nucleation scenarios. The
resulting interior MT distributions are simply those obtained by
\textquotedblleft cropping\textquotedblright\ the isotropic distribution to
the region defined by the cell, while the distribution on the boundary is
the radial projection of the distribution outside the cell onto the surface.

With the results on the MT number distributions, we are finally in a position to
give the sought-after length distribution: 
\end{subequations}
\begin{subequations}
\label{eq:l_density}
\begin{align}
l(\varphi )& =m\frac{\bar{t}\left( \bar{l}-(\bar{l}+l_{\mathrm{b}}(\varphi
))F_{\mathrm{b}}(\varphi )\right) +t_{\mathrm{b}}l_{\mathrm{b}}(\varphi )F_{\mathrm{b}}(\varphi )}{%
t_{\mathrm{tot}}(\varphi )} \\
l(\varphi )& =\frac{m}{2\pi }\frac{\bar{t}\left( \bar{l}-(\bar{l}%
+l_{\mathrm{b}}(\varphi ))F_{\mathrm{b}}(\varphi )\right) +t_{\mathrm{b}}l_{\mathrm{b}}(\varphi )F_{\mathrm{b}}(\varphi )%
}{\tilde{t}_{\mathrm{tot}}}
\end{align}

\subsubsection{Predictions from the theory}
\label{sec:M0_predictions}
We now use the theory derived above to map out the behavior of model M0. As
our focus throughout is on the influence of the geometry and the boundary
interactions, we a priori fix the relevant dynamical parameters of the MTs
to a set of generic ones chosen on the basis of experimental data, and shown
in Table \ref{tbl:MTdynpar}. First, and foremost, these parameters fix the
mean length the MTs would have in the absence of any confining boundary
given by Eq.\ (\ref{eq:l_bar}) to $\bar{l}=$\SI{2.54}{\micro\metre}.

\begin{table}[ptb]
\setlength{\tabcolsep}{22pt}
\par
\begin{center}
{\small 
\begin{tabular}{p{3cm}p{1cm}p{1.6cm}p{1cm}}
\hline
Parameter & Symbol & Value & Reference \\ \hline
Growth speed & $v_{+}$ & \SI{0.018}{\micro\metre\per\second} & \cite{Su2013} \\ 
Shrinkage speed & $v_{-}$ & \SI{0.040}{\micro\metre\per\second} & \cite{Su2013} \\ 
Nucleation rate & $r_{\mathrm{n}}$ & \SI{0.05}{\per\second} & \cite{Vogel2001} \\ 
Catastrophe rate & $r_{+}$ & \SI{0.0078}{\per\second} & \cite{Su2013} \\ 
Rescue rate & $r_{-}$ & \SI{0.0016}{\per\second} & \cite{Su2013} \\ \hline
\end{tabular}
}
\end{center}
\caption{Basic dynamical parameters of MTs used throughout.\label{tbl:MTdynpar}}
\end{table}

The first question we address is the influence of the overall scale of the
cell, as compared to $\bar{l}$, on the organisation of the MTs. Since the interactions of the MTs with the boundary do not depend on the location on the boundary, we expect the MT distributions to follow the biaxial symmetry of the cell. We therefore compare the
value of the bipolar order parameter $S_2$ for different values of the
boundary residence time $t_b$ at a fixed aspect ratio $b/a=4$ for different
absolute sizes of the cell, chosen such that three distinct relevant cases are
covered: $a<b<\bar{l}$, $a <\bar{l}<b$ and $\bar{l}<a<b$. 

The main takeaway of the results shown in Fig.~\ref{fig:M0_size}(a) is that in the random
nucleation scenario the order parameter $S_2$ varies more strongly as a function of the residence time
than in the homogeneous nucleation scenario. This is readily understood as in
the random scenario the MTs can be effectively redistributed over the possible
orientations if they are ``sequestered'' at the boundary, whereas in the
homogeneous scenario fixed numbers of MTs are apportioned to each interval of
angles. Another striking result is that in the largest cells, where MTs will
hardly ever reach the longitudinal poles, $S_2$ will become negative in both
scenarios as the distribution is dominated by MTs captured at the boundary
on the transverse short axis. Since we are explicitly interested in the
competition between the short and the long axis of the cell, we now choose
to fix to the short semi-axis to $a=$\SI{1}{\micro\metre}, and consider three cases in
which the probability of MTs reaching the pole in the longitudinal direction
is high ($b<\bar{l}$), average ($b=\bar{l}$), and low ($b>\bar{l}$),
respectively. The results are given in Fig.~\ref{fig:M0_size}(b) showing once
again that the random nucleation scenario displays the largest sensitivity
to the residence time at the boundary.

\begin{figure}[h]
\centering
\subfigure[]{\includegraphics[width=0.485\textwidth]{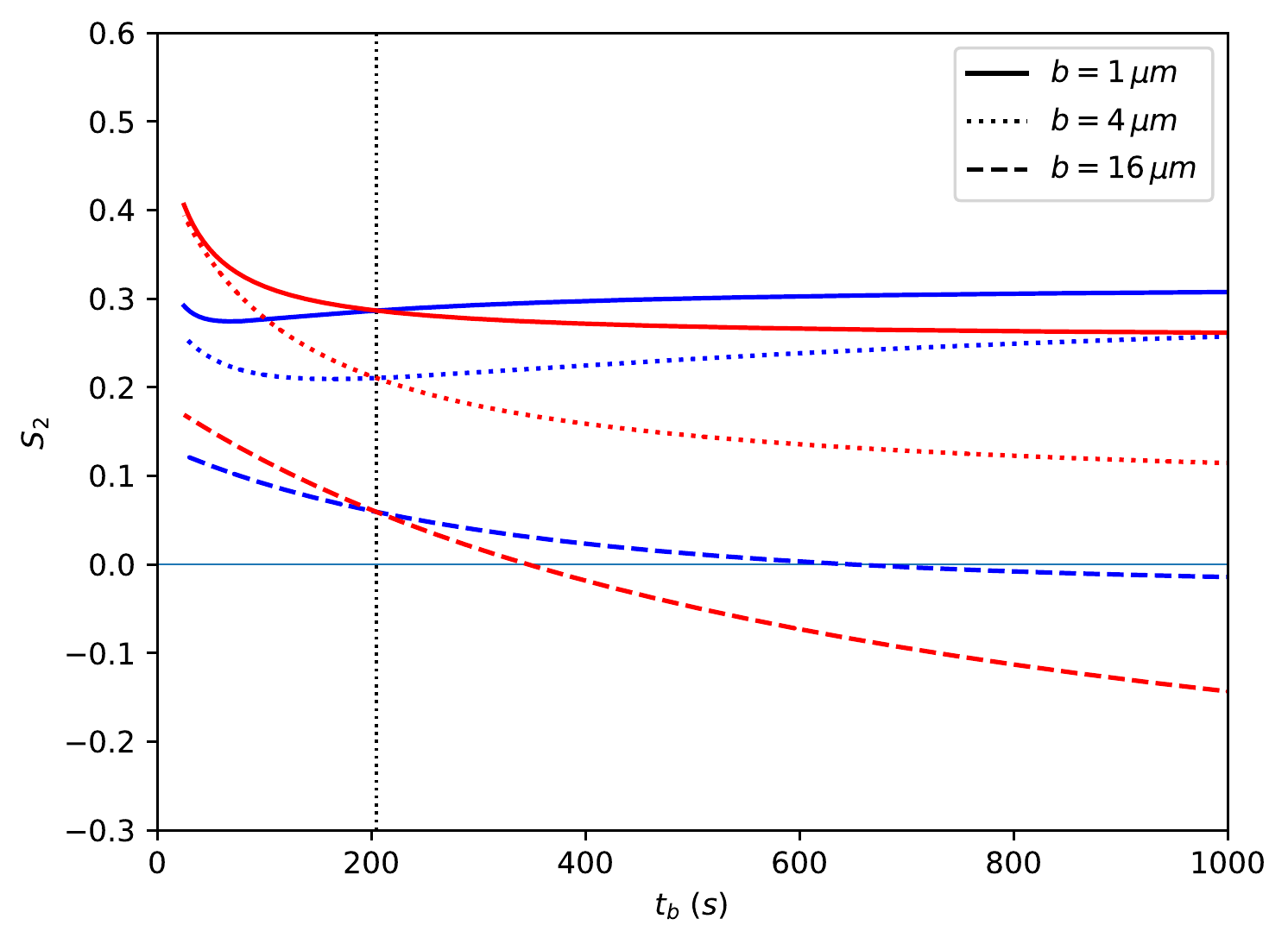}} 
\subfigure[]{\includegraphics[width=0.485\textwidth]{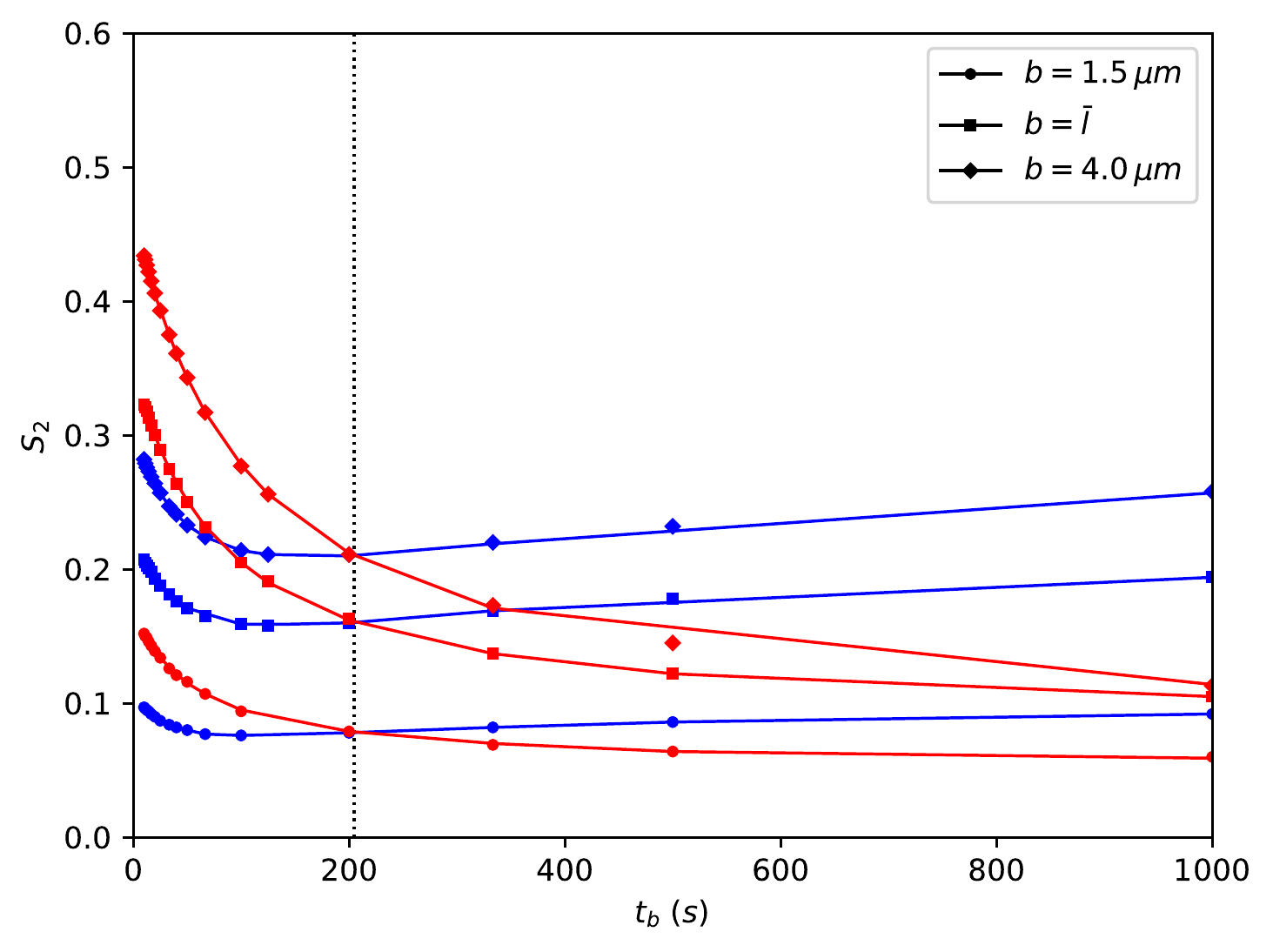}}
\caption{\textbf{Model M0}: Order parameter $S_2$ as a function of the
unbinding rate $r_{\mathrm{u}}$, in both nucleation scenarios: homogeneous (blue) and
random (red). The vertical line at $t_b=$\SI{204.8}{\second} indicates the special
value of the boundary residence time at which the difference between the
nucleation scenarios vanishes. (a) For a fixed aspect ratio $b/a=4$. Cases:
(i) $a=$\SI{0.25}{\micro\meter}, $b=$\SI{1}{\micro\meter} (solid), (ii) $a=$\SI{1} {\micro\meter}, $b=$\SI{4}{\micro\meter} (dotted), (iii) $a=$\SI{4} {\micro\meter}, $b=$\SI{16} {\micro\meter} (dashed). (b) For different aspect ratios $b/a$ setting $a=$\SI{1}{\micro\meter}
and cases (i) $b=$\SI{1.5}{\micro\meter}, (ii) $b=\bar{l}=$\SI{2.54} {\micro\meter} and
(iii) $b=$\SI{4}{\micro\meter}. The symbols indicate the values obtained by
simulations. Error bars in the simulations are smaller than the symbols. \label{fig:M0_size}}
\end{figure}

On the basis of these results and the pragmatic need to reduce the parameter
space addressed by our study, we now make the following choices which we
will apply throughout the rest of the study: (i) as it appears most sensitive probe of changes in the MT organisation 
due to changes in the geometry and/or the boundary residence time, we adopt the random
nucleation scenario, (ii) to explicitly address the competition between the
two axes of the cell, we adopt $a=$\SI{1}{\micro\meter} ensuring that the cell boundary
in the transverse direction is readily accessible to MTs, and (iii) as $t_{\mathrm{b}} =$\SI{100}{\second}
lies in the middle of the regime of largest sensitivity, we choose this
value of the boundary residence time to be our baseline, close to the
experimentally reported value of $\approx$\SI{90}{\second} in fission yeast \cite{Tran2001}.

\subsubsection{Comparison with simulations}

In order to set up the core algorithm which will be used to simulate the MT dynamics in the rest of the study, we perform
standard fixed time step stochastic simulations of model M0.  These simulations are then validated against the analytical predictions of the previous section. 

In the simulations, individual MTs are modelled as objects in one of the possible
states \textsc{dormant}, \textsc{growing}, \textsc{shrinking} and \textsc{%
bound}. At each time step the probability of transitioning to another state
is calculated and sampled. If the MT remains in its state its length is
updated as appropriate. The possible transitions are \textsc{dormant} $%
\rightarrow$ \textsc{growing}, with rate $r_{\mathrm{n}}$, \textsc{growing} $%
\rightarrow $ \textsc{bound}, which occurs whenever the length of the MT is
equal to the distance between MT and cell boundary in the direction in which
it is growing and \textsc{bound} $\rightarrow$ \textsc{shrinking}, with rate 
$r_{\mathrm{u}}$, \textsc{growing} $\rightarrow$ \textsc{shrinking} (catastrophes),
with rate $r_{+}$, \textsc{shrinking} $\rightarrow$ \textsc{growing}
(rescues), with rate $r_{-}$ and finally, \textsc{shrinking} $\rightarrow$ 
\textsc{dormant}, which occurs whenever a shrinking MT hits zero length. In
the homogeneous nucleation scenario, each MT is assigned a fixed angle $%
\varphi _{\mathrm{m}}=m\,2\pi /M,\, m=0,1,\ldots,M-1$. In the random scenario, a random angle
is chosen upon a nucleation event. Here we chose to simulate $M=1000$ MTs
and use a time step of $\Delta t=$\SI{0.5}{\second}. The dynamical parameters of the MTs
are the ones given in Table \ref{tbl:MTdynpar} above. Here, as in the rest of the study, we use the order parameters defined in Section \ref{sec:geom_dynamics} as reporters on the global organisation of the MTs. In Appendix \ref{app:distrib} we show a single representative comparison between the distribution function $l(\varphi)$ obtained by solving Eq.\ (\ref{eq:l_density}) and the distribution measured in simulations, from which the order parameters are then derived.

\subsection{Model MS: Force generation and boundary sliding}
\label{sec:MS}
In this second model we take into account that, due to continued
polymerization, a MT stalled at the cell boundary exerts a force in the direction
in which it is oriented \cite{Dogterom2005}. In a non-spherical cell this force
generically has a component tangential to the surface, which can cause the
MTs to slide along the surface \cite{ma2014}. At the same time, one expects
that the growth speed as well as the catastrophe rate of bound MT are
influenced by the loading force \cite{Janson2003}. Here we will take all
these effects into account using a recently developed model of dynamic force
generation, which is parametrized using data on yeast cells \cite{Teapal2021ForcedCells}. A generic
friction parameter then controls the degree to which sliding contributes to
the overall MT organization. This model is illustrated in Fig.\ \ref{fig:model_MS}

\begin{figure}[h]
\centerline{\includegraphics[width=0.75\textwidth]{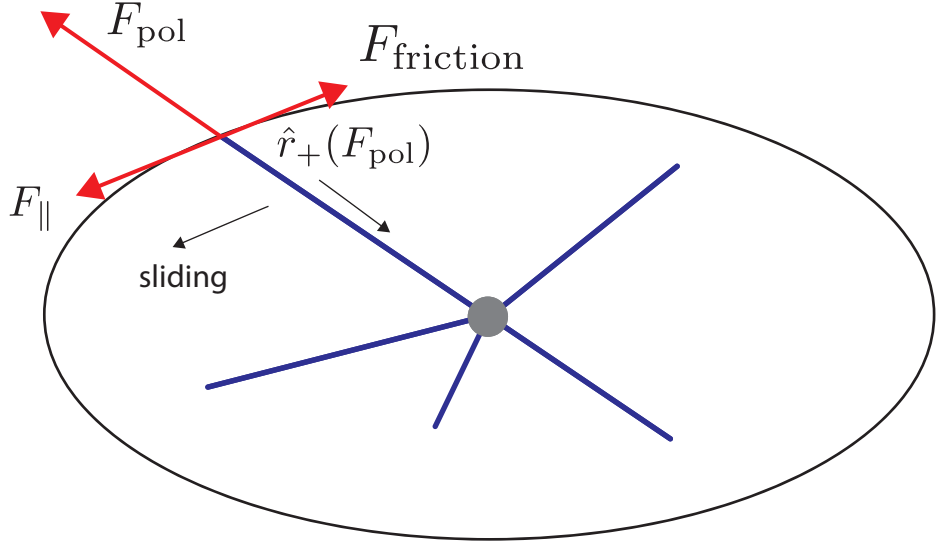}}
\caption{\textbf{Model MS}: Tangential component $F_{\parallel}$ of the
polymerization force $F_{\text{pol}}$ counteracted by the friction force $F_{%
\text{friction}}$ causes bound MTs to slide along the cell boundary. Their
residence time at the boundary is influenced by their degree of force
loading, through a force-dependent catastrophe rate $\hat{r}_{+}(F_{\text{pol%
}})$. \label{fig:model_MS}}
\end{figure}

\subsubsection{Dynamic force generation mechanism}

We adopt the force generation model described in \cite{Teapal2021ForcedCells}. This model is based on the phenomenological notion of ``stored length'', which is built up when the MT continues to grow after coming into contact
with the boundary. That MTs are able to grow due to thermal fluctuations, in
spite of the fact that they are in contact with a boundary, is a key ingredient of the
standard model Brownian-ratchet of polymerization forces \cite%
{Peskin1993CellularRatchet}. The stored length, defined as the difference between the length of MT and the distance between the nucleation point and the point of contact with the boundary, is interpreted as the
source of a linear ``expansion'' force countering the compression at the
boundary given by 
\end{subequations}
\begin{equation}  \label{eq:Fpol}
F_{\text{pol}}(l,L)=k(l-L),
\end{equation}
where $l$ is the length of the MT, $L$ the distance between MTOC and the
cell boundary and $k$ an effective compression modulus which governs the
magnitude of the generated force. We next assume that the microscopic
dynamics of the MT, involving the addition and removal of individual tubulin
subunits, is fast compared to the growth process, and that the off-rate by which tubulin subunits detach from the MTis
small with respect to the on-rate by which subunits attach to the MT. Under these assumptions, the growth speed
quasi-statically decreases as \cite{Dogterom1997} 
\begin{equation}  \label{eq:v_hat}
\hat{v}_{+}(F_{\text{pol}})=v_{+} e^{-\beta d F_{\text{pol}}},
\end{equation}
where $\beta=(k_B T)^{-1}$ is the inverse temperature in units of the
Boltzmann constant and $d$ is the microscopic step size of the growth
process. This parameter can be determined phenomenologically on the basis of
measured force-velocity relations. The value suggested by Foethke et al.\ 
\cite{Foethke2009}, in the context of a similar model, is presented as $%
\beta d=1/f_{\mathrm{s}}$ with the characteristic force $f_{\mathrm{s}}=$\SI{1.67}{\pico\newton} determining the
sensitivity of the MT growth to the opposing force. Throughout we will
denote quantities in the force-loaded state by a hat $(\hat{.})$.

To model the effect that the catastrophe rate should increase when in
contact with the boundary, the assumption is made that the linear relation
observed between growth speed and mean time to catastrophe for freely
growing MTs \cite{Janson2003} also holds instantaneously for loaded MTs.
This implies that 
\begin{equation}
\hat{t}_{+}(F_{\text{pol}})=\frac{1}{\hat{r}_{+}(F_{\text{pol}})}\propto 
\hat{v}_{+}(F_{\text{pol}}),
\end{equation}
which yields 
\begin{equation}  \label{eq:r_hat}
\hat{r}_{+}(F_{\text{pol}})=r_{+} e^{F_{\text{pol}}/f_{\mathrm{s}}}.
\end{equation}
Moreover, we assume that in the loaded state, $l > L$, no rescues are
possible, so that once a catastrophe occurs in this regime the MT will
shrink to the unloaded state $l \le L$ and that the shrinkage speed is
unaffected by the loading.

For non-spherical cells, the growth force is generically not perpendicular
to the boundary and so has a tangential component $F_{\parallel}$. This
component of the force is counteracted by the net damping force experienced
by the sliding motion of the MT along the boundary with velocity $%
v_{\parallel}$. Velocity and force are thus connected by the viscous
equation of motion 
\begin{equation}
F_{\parallel}=-\xi v_{\parallel},
\end{equation}
where $\xi$ is the effective drag coefficient.

\subsubsection{Implementation}

\label{sec:sim_MS} In order to implement force production and sliding into
our stochastic simulations, we replace the \textsc{bound} state of model M0,
by the state \textsc{pushing}. In the latter state the MT grows with speed $%
\hat{v}_{+}(F_{\text{pol}})$ given by Eq.\ (\ref{eq:v_hat}) and experiences
a catastrophe rate given by $\hat{r}_{+}(F_{\text{pol}})$ given by Eq.\ (\ref%
{eq:r_hat}), where the polymerization is found from the current length and
orientation through Eq.\ (\ref{eq:Fpol}). If at the end of a time step the
MT remains in the \textsc{pushing} state, the tangential force $F_{\parallel
}$ it experiences is calculated using Eq.\ (\ref{eq:Fpol}) by projecting
onto the tangent line to the elliptical boundary at the point of contact
determined by its current orientation $\varphi $. The MT is then rigidly
rotated over an angle $\Delta \varphi =F_{\parallel }/(\xi L(\varphi
))\Delta t$. Ignoring catastrophes, the stable points of this rotation are
the poles of the ellipse on the long axis, where the tangential component of
the forces disappears. By the same token, generically $L(\varphi +\Delta
\varphi )<L(\varphi )$ so that this motion also relaxes the magnitude of the
driving force.

In order to facilitate comparison between the models MS and M0, we must
choose a suitable value for the effective modulus $k$. We do this by
requiring that in the absence of sliding, the mean time until
catastrophe of a pushing MT equals the mean residence time set by the
reference unbinding rate $r_{\mathrm{u}}$ discussed in Section \ref{sec:M0_predictions}. To be fully precise, the residence time should also include the time it takes a loaded MT to shrink to the unloaded state, but given that the shrinking speed is significantly larger than the growing speed, this would only amount to a small correction. The mean time to catastrophe in the force production model works out as \cite%
{Teapal2021ForcedCells} 
\begin{equation}  \label{eq:tau_c}
\langle\tau_{\mathrm{c}}\rangle=\frac{1}{r_{+}}\sqrt{\pi}\sqrt{r_{+}\Phi}e^{r_{+}\Phi} %
\erfc(\sqrt{r_{+}\Phi}),
\end{equation}
where 
\begin{equation}  \label{eq:Phi}
\Phi=\frac{f_{\mathrm{s}}}{2 k v_{+}},
\end{equation}
and $\erfc{}$ is the complementary error function (see Ref.\ \cite{Gradshteyn2007TableEdition}, Table entry 8.250.4),
and $r_{+}$ and $v_{+}$ are the force-free values of the catastrophe rate
and the growth speed respectively. We now adjust $k$ to achieve $%
\langle\tau_{\mathrm{c}}\rangle=1/r_{\mathrm{u}}$ for the reference case $r_{\mathrm{u}}=$\SI{0.01}{\per\second}, which
yields $k=$\SI{0.3}{\pico\newton\per\micro\metre}.

\subsubsection{Simulation results}

\begin{figure}[h]
\centerline{\includegraphics[width=0.7\textwidth]{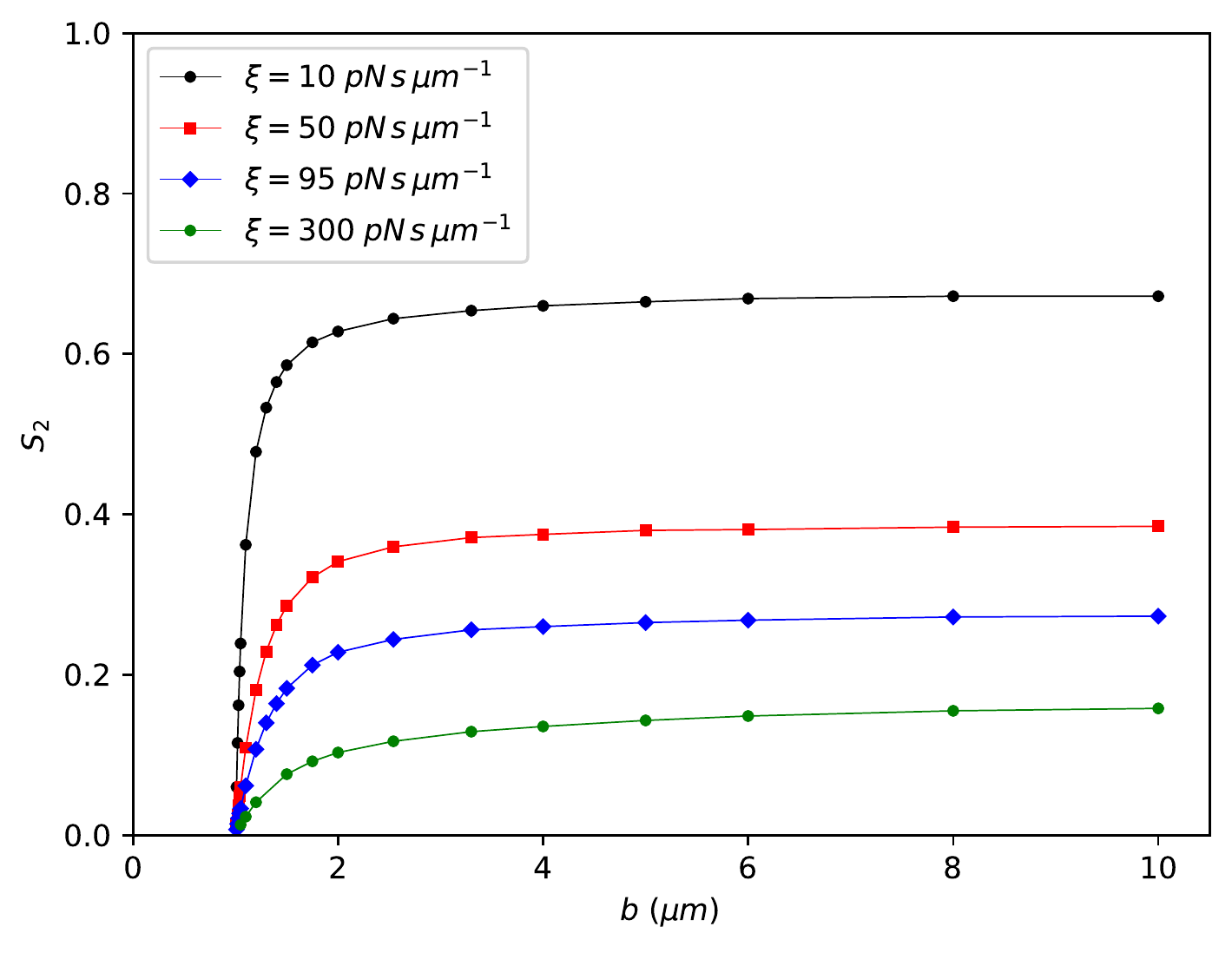}}
\caption{\textbf{Model MS}: Order parameter $S_2$ for different aspect
ratios by changing the long semi-axis $b$ ($a=1 \mu m$) for different values of
the sliding drag coefficient $\xi$. Lines through the data points are guides to the eye.}
\label{fig:S2_b_slide}
\end{figure}

We apply our algorithm to different aspect ratios $b/a$ of the cell and to different values of the sliding drag coefficients $\xi$. The simulation results for the order parameter $S_2$ are shown in Fig.\ \ref{fig:S2_b_slide}. In all cases, we see that the sliding mechanism leads to robust biaxial order dominant along the longitudinal axis ($S_2>0$), with the degree of ordering increasing with decreasing friction, and reaching values significantly above those achieved in the reference Model M0 (cf.\ Figure \ref{fig:M0_size}(b)).

\subsection{Model MP: Molecular polarization mechanism}

\label{sec:MP} 
\begin{figure}[h]
\centerline{\includegraphics[width=0.75\textwidth]{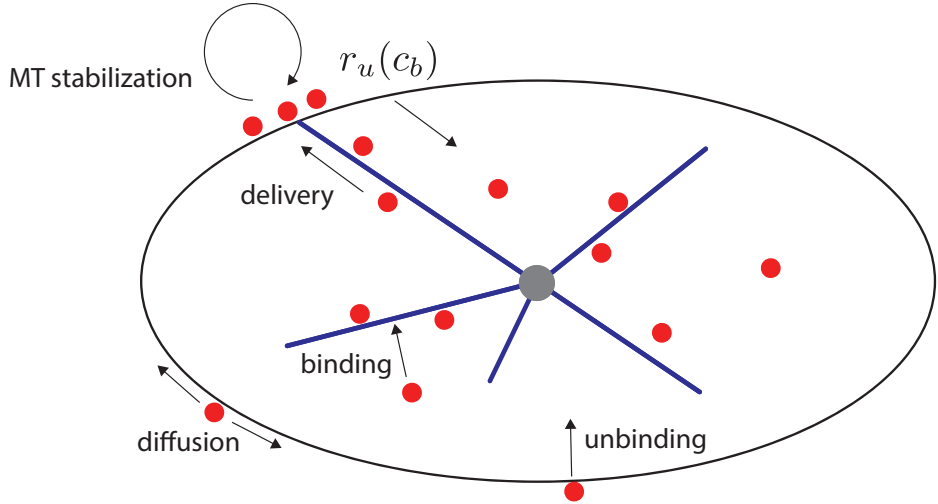}}
\caption{\textbf{Model MP}: The effector species PF (red circles) binds to
MTs over which they are transported. Bound MTs deliver PFs to the membrane,
where they diffuse until they unbind to return to the cell interior. The
residence time of bound MTs depends on the local density of PFs $c_{\mathrm{b}}$ in the
membrane through the unbinding rate $r_{\mathrm{u}}(c_{\mathrm{b}})$.}
\label{fig:model_MP}
\end{figure}

In this third model we add a MT-configuration dependent
polarization mechanism to the basic model M0. The main additional ingredient
of this model is the presence of a pool of effector proteins, which we dub
polarity factors (henceforth PFs). These PFs start out cytosolic, i.e.\ in
the cell interior, where they diffuse and bind to MTs. Bound PFs are
transported in the plus-end direction along their host MTs towards the cell
periphery. If their host MT is at the cell boundary, they can be delivered
into the membrane, in which they diffuse until they unbind and recycle into
the cell interior. The key assumption of our polarization mechanism is that
the residence time of the bound MTs depends on the local density of
membrane-bound PFs setting up a positive feedback loop: the higher the local
PF density, the longer a MT remains bound, the more PFs it delivers. As the
total pool of PFs is finite, this also causes a global depletion effect,
which represses the polarizing propensity of MTs in other parts of the cell.
Conceptually this model thus belongs to the generic class of
activator-depletion models (see \cite{jilkine2011} for a general overview),
but distinguishes itself by employing the non-diffusible MTs as a mediator
species. The model is schematically illustrated in Fig.~\ref{fig:model_MP}.

\subsubsection{Formalism and implementation}

The model adopts the formalism developed in \cite{Foteinopoulos2017APolarity}%
, where it was applied in a spherical cell geometry. Conservation of PFs
implies that at any time 
\begin{equation}
C=C_{\mathrm{f}}(t)+C_{\mathrm{m}}(t)+C_{\mathrm{b}}(t),
\end{equation}%
where $C$ is the total number of PFs, $C_{\mathrm{f}}(t)$ the number of free PFs in
the interior, $C_{\mathrm{m}}(t)$ the number of PFs bound to MTs and $C_{\mathrm{b}}(t)$ the
number of PFs bound to the membrane. We assume the diffusion of the PFs in
the cell interior to be very fast, so that their instantaneous distribution
is spatially homogeneous, and that the kinetics of binding and unbinding to
the MTs is so fast that an instantaneous binding-equilibrium is established,
allowing the linear density of PFs bound to MTs to be given by 
\begin{equation}
c_{\mathrm{m}}\left( t\right) =\frac{C_{\mathrm{m}}\left( t\right) }{L\left( t\right) }=\frac{1%
}{L\left( t\right) +L_{\frac{1}{2}}}\left( C-C_{\mathrm{b}}(t)\right) ,  \label{eq:cm}
\end{equation}%
where $L$ is the total length of all MTs in the system, and $L_{\frac{1}{2}}$
a parameter that sets the affinity of the PFs for binding to the MTs.
Calling the constant transport speed of PFs bound to MTs $v_{\mathrm{m}}$, each
membrane-bound MT delivers $v_{\mathrm{m}}c_{\mathrm{m}}(t)\Delta t$ PFs to the membrane per
time step. Once in the membrane the PFs perform a standard diffusion and can
unbind at a rate $k_{\mathrm{u}}$, in which case they return to the interior pool.
The coupling between membrane-bound PFs and membrane-bound MTs is
implemented by the non-linear dose response function, which governs the MT
unbinding rate as a function of the local PFs density $c_{\mathrm{b}}$ 
\begin{equation}
r_{\mathrm{u}}(c_{\mathrm{b}})=\left( r_{\mathrm{u}}(0)-r_{\mathrm{u}}(\infty )\right) \frac{1}{1+\left( \frac{%
c_{\mathrm{b}}}{c_{\ast }}\right) ^{p}}+r_{\mathrm{u}}(\infty ), \label{eq:ru_P}
\end{equation}%
where $r_{\mathrm{u}}(0)$ is the MT-unbinding rate in the absence of PFs, $%
r_{\mathrm{u}}(\infty )$ the MT-unbinding rate at PF oversaturation, $c_{\ast }$ a
cross-over density and $p$ a Hill-coefficient, which governs the steepness
of the cross-over between the low- and high-density regime.

In the simulations, the diffusion of the PFs is implemented as a fixed time
step continuous space Brownian motion obtained by sampling from the
appropriate Gaussian propagator. The local density is evaluated by binning
the PFs in the boundary, with an additional discrete noise suppressing
averaging over a local neighborhood. For further details the reader is
referred to \cite{Foteinopoulos2017APolarity}. The values of the additional parameters
used are shown in Table \ref{tab:polmodpar}.
\begin{table}[h]
    \centering
    \begin{tabular}{lcl}
    \hline
        Parameter &  Symbol & Value\\ \hline
        Binding affinity PFs to MTs & $L_{\frac{1}{2}}$ & \SI{150}{\micro\metre} \\
        Transport speed PFs along MTs & $v_m$ & \SI{0.81}{\micro\metre\per\second} \\
        Base MT unbinding rate & $r_{\mathrm{u}}(0)$ & \SI{0.01}{\per\second} \\
        MT unbinding rate at PF saturation & $r_{\mathrm{u}}(\infty)$ & \SI{0.001}{\per\second} \\
        Hill coefficient dose-response curve & $p$ & $5$ \\
        Cross-over density dose-response curve & $c_{*}$ & $20 \text{bin}^{-1}$ \\
        Diffusion coefficient PFs & $D$ & \SI{0.035}{\micro\metre\squared\per\second} \\
        PF unbinding rate & $k_u$ & \SI{0.07}{\per\second} \\
        \hline
    \end{tabular}
    \caption{Parameters used in the implementation of the polarisation mechanism.}
    \label{tab:polmodpar}
\end{table}

\subsubsection{Simulation results}
\begin{figure}[h]
\centering
\subfigure[]{\includegraphics[width=0.485\textwidth]{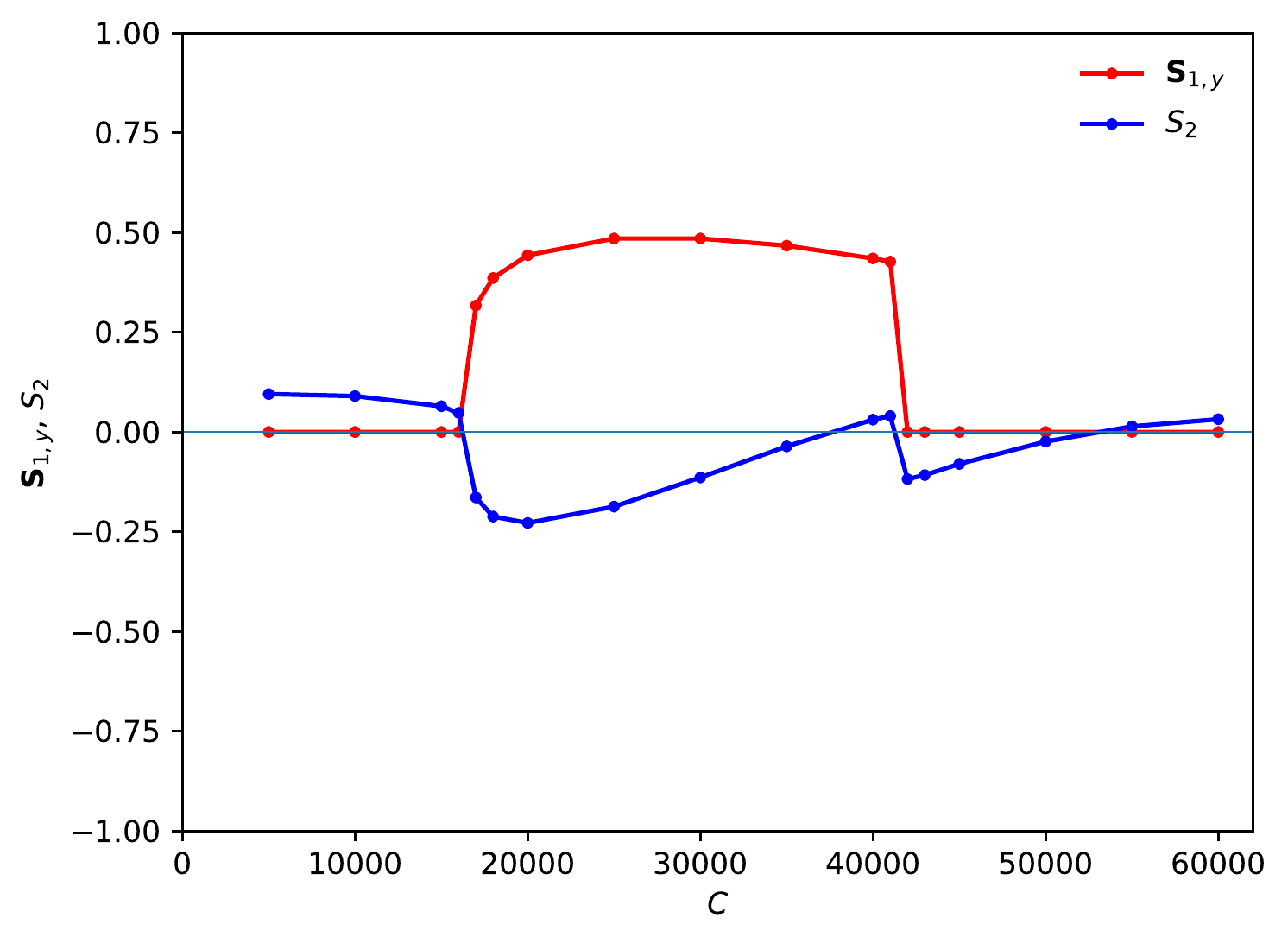}} 
\subfigure[]{\includegraphics[width=0.485\textwidth]{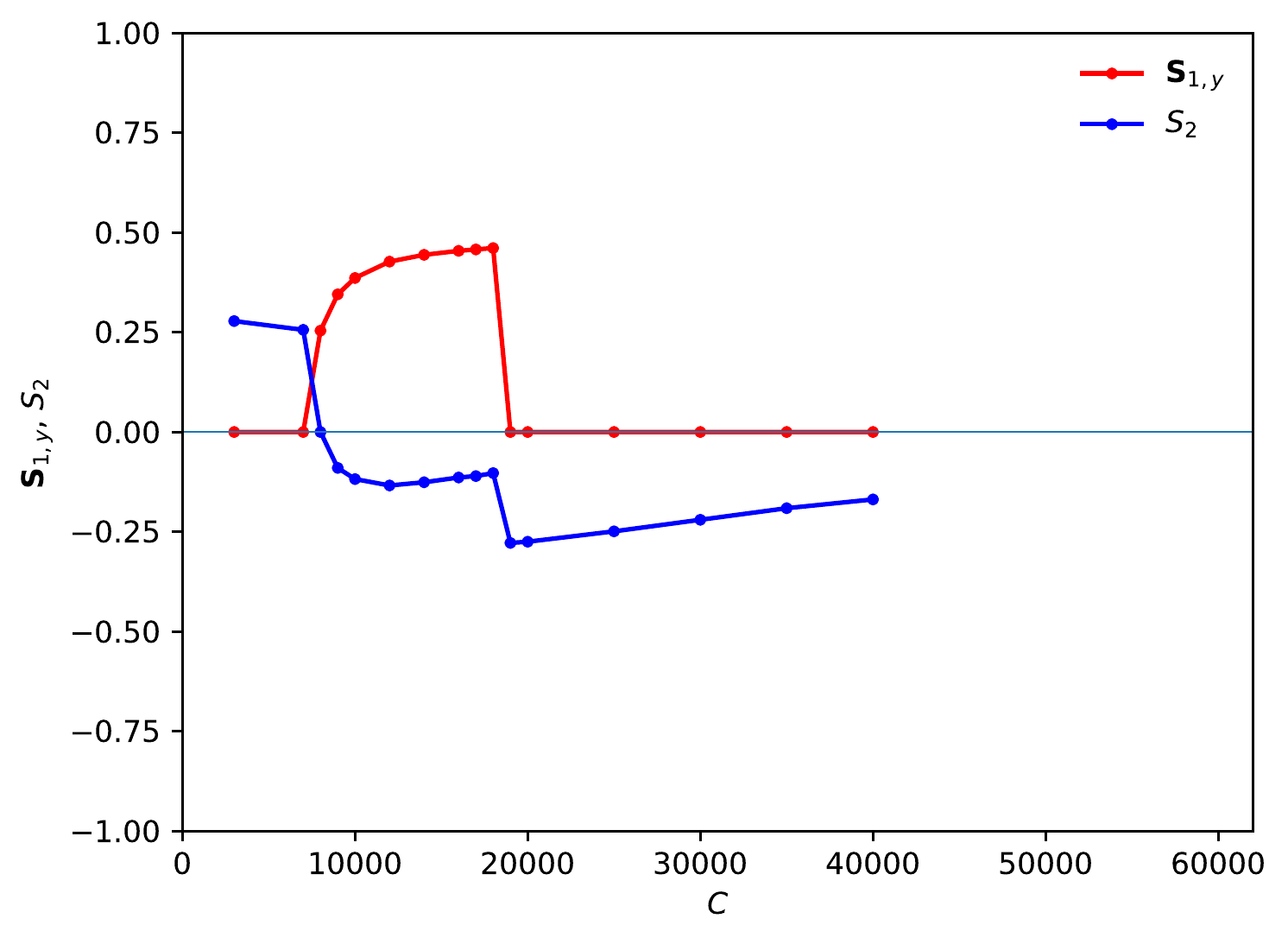}}
\caption{\textbf{Model MP}: Order parameters $|\mathbf{S}_{1,y}|$ and $S_2$ as a function
of the total number of polarity factors in the system, for (a) $b=$\SI{1.5}{\micro\metre}
and (b) $b=$\SI{4}{\micro\metre}.}
\label{fig:MP_order}
\end{figure}

We simulated model MP for cellular geometries with two different aspect ratios: a less elongated, and hence more nearly circular, case with $b/a=1.5$ and a more elongated case $b/a=4$. The results are shown in Figs.\ \ref{fig:MP_order}(a) and \ref{fig:MP_order}(b), respectively. We see that in both cases there is a range of values for $C$, the total number of PFs in the system, for which polarization is observed. Since the number of MTs in contact with the boundary is largest along the short, transverse axis of the cell the polarization occurs along this axis. We therefore only plotted the component $|\mathbf{S}_{1,y}|$ of the vectorial order parameter $\mathbf{S}_1$. The absolute value is taken for convenience, as by reflection symmetry in the $x$-axis, the polarization in the $-y$ direction is as likely as in the $+y$-direction. 

The most striking result is observed for the non-polarized states in the more elongated cell (Fig.\ \ref{fig:MP_order}(b)). For low values of $C$, where the polarization mechanism has not yet kicked-in, the system responds to the geometry similarly to the reference Model M0 (cf.\ Fig. \ref{fig:M0_size}(b)), i.e.\ with a slight preference for longitudinal biaxial order ($S_2>0$). However, at high values of $C$ when the polarization mechanism is no longer effective due to oversaturation, the system actually retains an `imprint' of the transverse polarization at intermediate values of $C$, by now settling on a transverse biaxial ordered state ($S_2<0$), with the major mass of the MT-length distribution oriented along the $y$-axis.

\subsection{Model MSP: Polarization mechanism and sliding}

\label{sec:MSP} 
\begin{figure}[h]
\centerline{\includegraphics[width=0.75\textwidth]{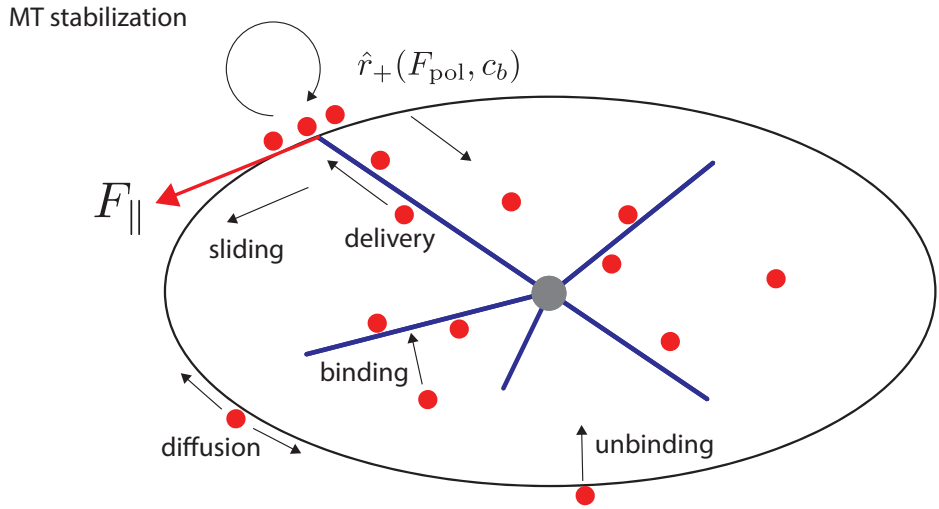}}
\caption{\textbf{Model MSP}: Model that combines the polarization mechanism
based on effector species PF (red dots) with polymerization-force induced
sliding. The residence time of bound MTs depends both on the polymerization
force and the local PF density through the catastrophe rate $\hat{r}_{+}(F_{%
\text{pol}},c_{\mathrm{b}})$}
\label{fig:model_MSP}
\end{figure}

In this final model we combine the sliding mechanism of
Model MS with the polarization mechanism of Model MP. We focus on the
interplay between the tendency of sliding to create a bipolar MT
organisation along the long axis of the cell, and the tendency of the
polarization mechanism to establish unipolar order along the short axis of
the cell. The model is schematically illustrated in Fig.\ \ref{fig:model_MSP}%
.

\subsubsection{Combining force production with the polarization mechanism}

In order to connect the force-production mechanism of Model MS to the
polarity-generation mechanism of Model MP, we chose to generalize the
procedure adopted in Section \ref{sec:sim_MS} to link model MS to M0 through
the mean-residence time. In this case we require that the mean time to
catastrophe due to the force production mechanism $\langle\tau_{\mathrm{c}}\rangle$
tracks the non-linear dependence of the unbinding rate on the PF density,
i.e. 
\begin{equation}
\langle\tau_{\mathrm{c}}\rangle(c_{\mathrm{b}})=\frac{1}{r_{\mathrm{u}}(c_{\mathrm{b}})},
\end{equation}
where $r_{\mathrm{u}}(c_{\mathrm{b}})$ is given by Eq.\ \ref{eq:ru_P}. Recalling Eqs.\ (\ref%
{eq:tau_c}) and (\ref{eq:Phi}), we in principle have some freedom in which
parameter to use to enforce this identity. We argue, however, that the most
natural one is the unloaded catastrophe rate $r_{+}$, which most directly
represents the intrinsic stability of the MT that is modulated by the
presence of the PFs. In practice, we therefore solved (c.f.\ Eq.\ (\ref{eq:tau_c}))
\begin{equation}  \label{eq:tau_c_cb}
\frac{1}{r_{\mathrm{u}}(c_{\mathrm{b}})} = \frac{1}{r_{+}(c_{\mathrm{b}})}\sqrt{\pi}\sqrt{r_{+}(c_{\mathrm{b}})\Phi}%
e^{r_{+}(c_{\mathrm{b}})\Phi}\erfc(\sqrt{r_{+}(c_{\mathrm{b}})\Phi})
\end{equation}
for $r_{+}(c_{\mathrm{b}})$ over a range of $c_{\mathrm{b}}$ values, constructing a look-up table
from which the appropriate value can be retrieved by interpolation when
needed in the simulations. The details of this procedure are given in
Appendix \ref{app:r+_MSP}.

\subsubsection{Simulation results}
\begin{figure}[h]
\centering
\subfigure[]{\includegraphics[width=0.485\textwidth]{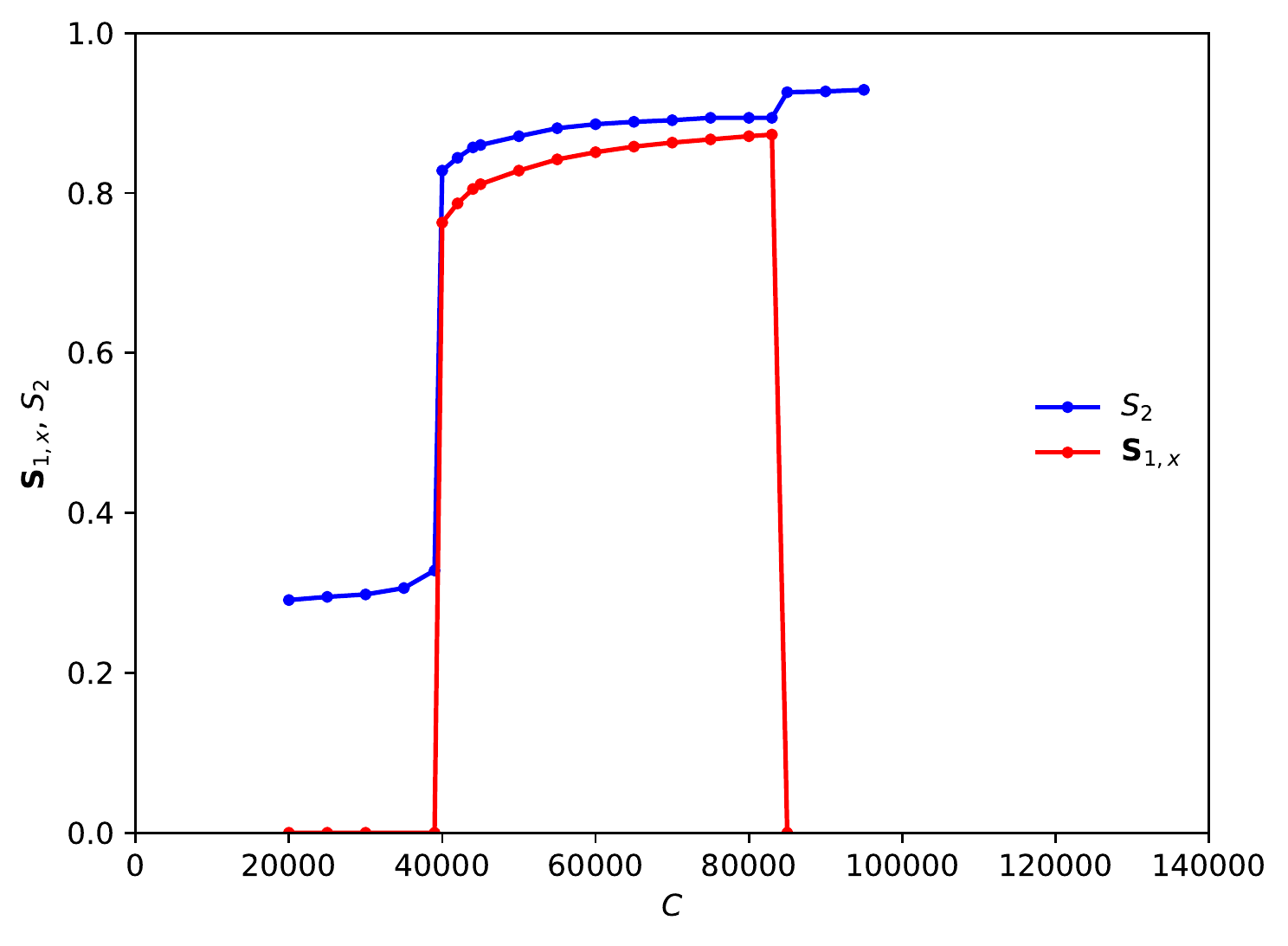}}
\subfigure[]{\includegraphics[width=0.485\textwidth]{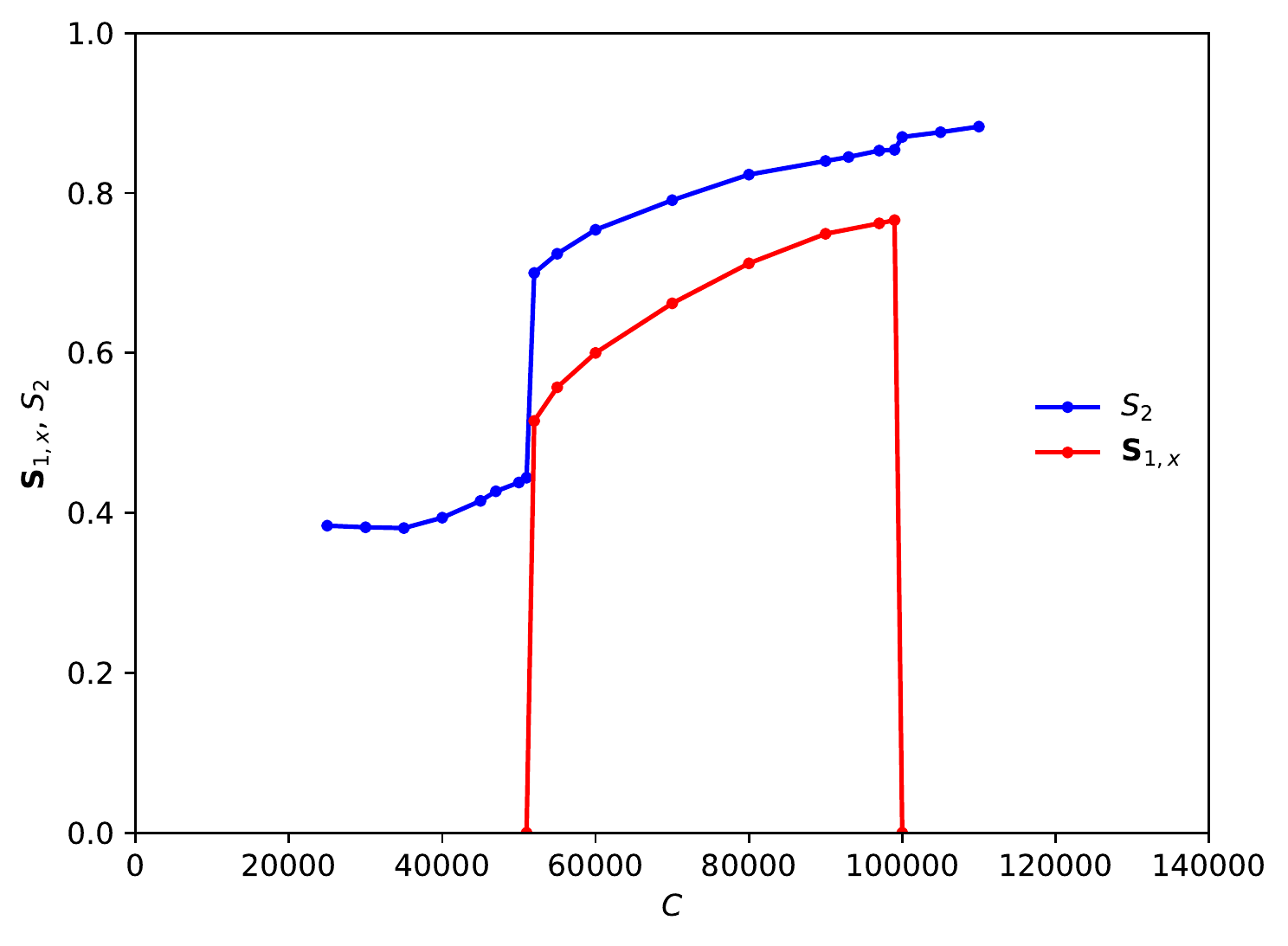}}
\caption{\textbf{Model MSP}: Order parameters $|\mathbf{S}_{1,x}|$ and $S_2$ as a function
of the total number of polarity factors in the system, for (a) $b=$\SI{1.5}{\micro\metre}
and (b) $b=$\SI{4}{\micro\metre}. \label{fig:MSP_order}}
\end{figure}

In Figure \ref{fig:MSP_order} we show the results of the simulations for two values of the aspect ratio of the cell plotting order parameters as function of $C$, the total number $C$ of PFs in the cell, for $\xi=$\SI{30}{\pico\newton\second\per\micro\metre}. Since in this case, due to the efficacy of the sliding mechanism, the number of MTs in contact with the boundary is largest along the longitudinal axis of the cell, polarization, when it occurs, is along this axis. Here, we therefore only plotted the component $|\mathbf{S}_{1,x}|$ of the vectorial order parameter $\mathbf{S}_1$, where, again for symmetry reasons, the absolute value is shown. In comparison with Model MP, the degree of polarization, which now piggybacks the intrinsic preference for longitudinal order already displayed in Model MS, is much more pronounced. At the same time, the impact of the polarization mechanism in the post-polarization high-$C$ regime on the degree of biaxial order is significantly higher than that achievable by geometry (c.f.\ Figure \ref{fig:M0_size}(b)) or sliding (c.f.\ Figure \ref{fig:S2_b_slide}) alone. 

\subsection{Toy-models}\label{sec:toymodel}
Considering the observed behavior of the models MP and MSP, we can readily discern the critical factor that distinguishes the two cases: whether or not the interaction of MTs with the boundary is dominated by the, by virtue of the innate exponential length distribution of the MTs, most accessible shortest distance in the geometry, i.e.\ the transverse one in the elliptical geometry. This suggests that the observed behavior can be recapitulated in the setting of a highly simplified toy-model that captures the essential ingredients at play. This model dispenses with the complexity due to to the continuous distribution of MT angles, and only considers a discrete number of directions. In the most generic case mimicking the behavior of model MP, we consider two opposing transverse orientations with a membrane at a distance $a$, and two opposing longitudinal orientations with the membrane at a distance $b$. Moreover, we remove any cross-talk due to PFs diffusing from one delivery point to another, effectively cutting the elliptical membrane into four disconnected sectors (see Fig.\ \ref{fig:toy_model}). The latter approximation is reasonable whenever the mean free diffusion length of PFs in the membrane is smaller than the distance between two delivery points, i.e.\ at quarter of the cell circumference. We will call this model TOY-MP. In case we are mimicking the behavior of model MSP, where, specifically when the friction coefficient is not too high, we know that sliding will focus the majority of MTs along the longitudinal a toy-model with just two orientations opposing longitudinal directions can be expected to be a fair approximation, which we dub TOY-MSP. The details of the construction of these two toy-models and our approach to numerically solve them are given in Appendix \ref{app:toy}.   

\begin{figure}[h]
\centering
\subfigure[]{\includegraphics[height = 10 cm]{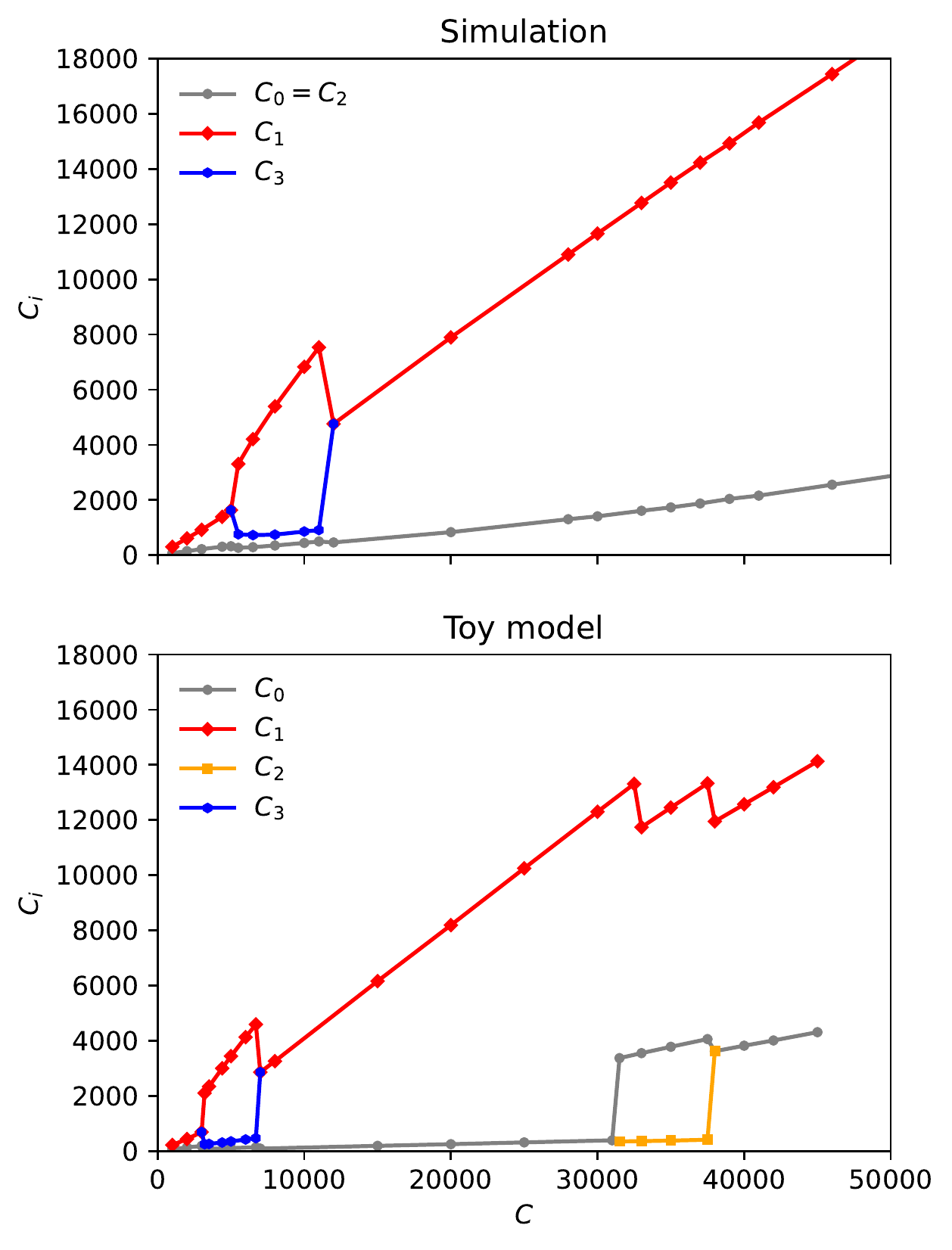}}
\subfigure[]{\includegraphics[height = 10 cm]{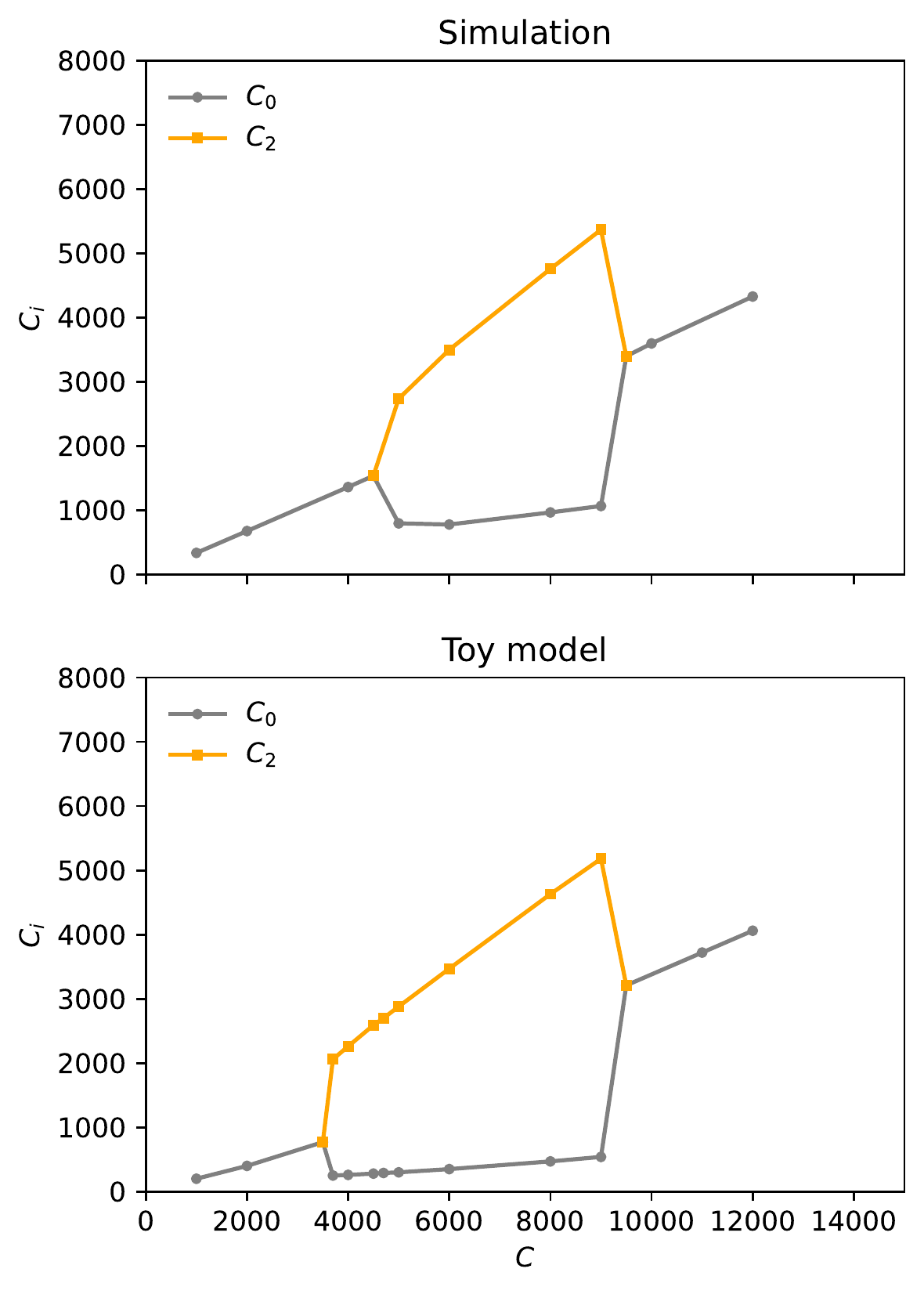}}
    \caption{(a) \textbf{Model TOY-MP}: Predictions for the (local) number of PFs $C_i$ in each of the four membrane sectors in each as a function of the total number of PFs $C$ of the toy model (below) compared to the full simulations (above). Here $C_{*}=1000$. (b) \textbf{Model TOY-MSP}: Predictions for the (local) numbers of PFs $C_i$ in the two halves of the membrane as a function of the total number of PFs $C$ of the toy model (below) compared to the full simulations with drag coefficient $\xi=$\SI{1}{\pico\newton\second\per\micro\metre} (above).}
    \label{fig:toy_MP_MSP}
\end{figure}

We can directly compare the results of the toy models to the full simulation, if we ensure that the fixed nucleation rate in the toy models is adjusted to the observed steady-state overall nucleation rate in the simulations. As observables, we take the total number of PFs in each membrane sector. The results for model TOY-MP are presented in Fig.\ \ref{fig:toy_MP_MSP}(a) and show that the toy-model indeed reproduces the transverse polarisation transition, albeit at a significantly lower number of total PFs. This stands to reason, as the PFs in the full model are far more dispersed over the membrane, in contrast to the toy-models where they are highly focused, and hence act more strongly to stabilize the bound MTs. Strikingly, the toy model also predicts a possible longitudinal polarisation transition, which occurs far beyond the point where the transverse polarisation has already disappeared due to local saturation of the polarisation mechanism. We did not observe such a transition in the full simulations, possibly because it occurs for a much higher number of PFs than we choose to simulate here. Model TOY-MSP, with just two directions, appears even to semi-quantitatively reproduce the full simulation data, as shown in  Fig.\ \ref{fig:toy_MP_MSP}(b).

\section{Discussion}\label{sec:discussion}
We have presented an exploration of the effects of non-spherical cell shape on the global distribution of MTs isotropically nucleated from a centrally located MTOC, a geometry that is a stylized version of the situation that pertains to a generic interphase eukaryotic cell, using four different models of increasing complexity.   The main effect is revealed in the simplest model M0 in which the MTs have a generic interaction with the cell boundary that causes them to stall for a variable amount of time. As in steady-state MTs have an exponential length distribution, they are much more likely to interact with the cell boundary in the transverse equatorial direction. This intrinsic orientational selection mechanism can be overruled if we allow MTs to slide driven by their intrinsic force-generating mechanism (Model MS), in which case the distribution can be reoriented towards the longitudinal direction. In both cases, the resulting distribution is bipolar with biaxial symmetry and conforms to the inversion symmetry of the cell shape. This strong coupling between the cell shape and MT distribution can itself be overridden by introducing an explicit polarisation mechanism. This mechanism is mediated by polarity factors that depend on MTs for their delivery to the cell membrane and in turn stabilize the bound MTs increasing their residence time at the boundary, hence setting up a positive feedback loop. This breaks the inversion symmetry and creates either a polarized distribution in the transverse direction (Model MP) or longitudinal direction (Model MSP). The various trade-offs involved are captured qualitatively, and in some cases even semi-quantitatively, by a very simple, and potentially extendable, toy-model that discretizes the orientations.

Obviously, the models presented here have a number of drastic (over)simplifications. The main ones concern the nature of the localization of the centrosome. In reality this structure is (i) eccentrically connected to the nuclear envelope, so that the nucleus will occlude a significant fraction of potential orientations for MTs, and (ii) not at a preordained location, but rather dynamically positioned. The latter effect has been studied extensively \cite{Laan2012,Laan2012a,pavin2012,ma2014} and likely involves the interplay between pushing forces (generated by the MTs themselves) and pulling forces (exerted by membrane-attached minus-end directed motor proteins). It is certainly feasible, albeit at the expense of a significant increase in complexity, to include both effects in a future version of these models.

Finally, the polarization mechanism with a single polarity factor employed here by construction leads to polar MT distributions. It is an interesting question from a fundamental point of view whether it is possible to create a polarization mechanism, likely involving at least two polarity factors, that supports biaxial polarization in the absence of the sliding mechanism. The latter could contribute to our understanding of the longitudinal biaxial MT organisation found, e.g. in fission yeast.

\begin{acknowledgments}
We thank Alex Cumberworth (AMOLF) for his critical reading of the manuscript. This work is part of the research programme of the Netherlands Organisation for Scientific Research (NWO) and was performed at the research institute AMOLF.
\end{acknowledgments}

\appendix

\section{Derivation state-steady equations model M0}

\label{app:model_M0} In order to derive the steady-state equations for the MT length distributions for model M0, we start from the time-dependent equations. For the active and bound MTs
these read 
\begin{align}
\partial _{t}m_{+}(l,\varphi ,t)& =-v_{+}\partial _{l}m_{+}(l,\varphi
,t)+r_{-}m_{-}(l,\varphi ,t)-r_{+}m_{+}(l,\varphi ,t)  \label{eq:m_plus} \\
\partial _{t}m_{-}(l,\varphi ,t)& =v_{-}\partial _{l}m_{-}(l,\varphi
,t)-r_{-}m_{-}(l,\varphi ,t)+r_{+}m_{+}(l,\varphi ,t)  \label{eq:m_minus} \\
\partial _{t}m_{\mathrm{b}}(\varphi ,t)& =-r_{\mathrm{u}}m_{\mathrm{b}}(\varphi
,t)+v_{+}m_{+}(l_{\mathrm{b}}(\varphi ),\varphi ,t).  \label{eq:m_bound}
\end{align}%
The equations for the dormant ones now depend on the chosen nucleation
scenario. For the homogeneous scenario there is a density $m_{0}(\varphi ,t)$
of dormant MTs per angle, while in the random scenario there is just a
single pool of dormant MTs $M_{0}(t)$. We thus have 
\begin{subequations}
\begin{align}
\partial _{t}m_{0}(\varphi ,t)& =-r_{\mathrm{n}}m_{0}(\varphi
,t)+v_{-}m_{-}(l=0,\varphi ,t)  \label{eq:m_0hom} \\
d_{t}M_{0}(t)& =-r_{\mathrm{n}}M_{0}(t)+v_{-}\int_{0}^{2\pi }\mathrm{d}\varphi
\,m_{-}(l=0,\varphi,t),  \label{eq:m_0ran}
\end{align}%
where throughout the $a$-sublabelled equations will refer to the homogeneous
case, and the $b$-sublabelled ones to the random case. These equations need
to be supplemented with boundary conditions. At the cell boundary these are 
\end{subequations}
\begin{equation}
r_{\mathrm{u}}m_{\mathrm{b}}(\varphi ,t)=v_{-}m_{-}(l_{\mathrm{b}}(\varphi ),\varphi ,t)
\label{eq:BCbt}
\end{equation}%
while at $l=0$ they again depend on the nucleation scenario 
\begin{subequations}
\label{eq:BC0t}
\begin{align}
v_{+}m_{+}(l=0,\varphi ,t)& =r_{\mathrm{n}}m_{0}(\varphi ,t)  \label{eq:BC0hom} \\
v_{+}m_{+}(l=0,\varphi ,t)& =\frac{1}{2\pi }r_{\mathrm{n}}M_{0}(t).  \label{eq:BC0ran}
\end{align}%
We now introduce a number of integrated quantities. First, the total number of growing, shrinking and \emph{active} MTs in a given direction in the interior, 
\end{subequations}
\begin{align}
m_{+}(\varphi ,t)& =\int_{0}^{l_{\mathrm{b}}(\varphi )}\mathrm{d}l\,m_{+}(l,\varphi ,t) \\
m_{-}(\varphi ,t)& =\int_{0}^{l_{\mathrm{b}}(\varphi )}\mathrm{d}l\,m_{-}(l,\varphi ,t) \\
m_{\mathrm{a}}(\varphi ,t)& =m_{+}(\varphi ,t)+m_{-}(\varphi ,t)
\end{align}%
and next the total number of active and bound MTs in the system
\begin{align}
M_{\mathrm{a}}(t)& =\int_{0}^{2\pi }\mathrm{d}\varphi \,m_{\mathrm{a}}(\varphi ,t) \\
M_{\mathrm{b}}(t)& =\int_{0}^{2\pi }\mathrm{d}\varphi \,m_{\mathrm{b}}(\varphi ,t). 
\end{align}%
 Adding Eqs.\ (\ref{eq:m_plus}), (\ref{eq:m_minus}), and integrating over the relevant lengths yields 
\begin{equation}
\partial _{t}m_{\mathrm{a}}(\varphi ,t)=\{v_{-}m_{-}(l_{\mathrm{b}}(\varphi ),\varphi
,t)-v_{+}m_{+}(l_{\mathrm{b}}(\varphi ),\varphi ,t)\}-\{v_{-}m_{-}(l=0,\varphi
,t)-v_{+}m_{+}(l=0,\varphi,t)\}.
\end{equation}%
Adding this identity to Eq.\ (\ref{eq:m_bound}) and taking into account Eq.\ (\ref%
{eq:BCb}) in the main text then gives 
\begin{equation}
\partial _{t}m_{\mathrm{a}}(\varphi ,t)+\partial _{t}m_{\mathrm{b}}(\varphi
,t)=v_{+}m_{+}(l=0,\varphi ,t)-v_{-}m_{-}(l=0,\varphi,t).  \label{eq:dtma+mb}
\end{equation}%
For the homogeneous nucleation scenario, this can immediately be combined
with Eq.\ (\ref{eq:BC0hom}) to yield 
\begin{equation}
\partial _{t}m_{0}(\varphi ,t)+\partial _{t}m_{\mathrm{a}}(\varphi ,t)+\partial
_{t}m_{\mathrm{b}}(\varphi ,t)=0,
\end{equation}%
while for the random scenario, we first need to integrate Eq.\ (\ref%
{eq:dtma+mb}) over all angles, and then combine with Eqs.\ (\ref{eq:m_0ran})
and (\ref{eq:BC0ran}) to get 
\begin{equation}
\frac{d}{dt}\left\{M_{0}(t)+M_{\mathrm{a}}(t)+M_{\mathrm{b}}(t)\right\}=0,
\end{equation}%
which given our definitions lead to the conservation equations 
\begin{subequations}
\begin{align}
m& =m_{0}(\varphi ,t)+m_{\mathrm{a}}(\varphi ,t)+m_{\mathrm{b}}(\varphi ,t) \\
M& =M_{0}(t)+M_{\mathrm{a}}(t)+M_{\mathrm{b}}(t).
\end{align}%
The steady-state equations used in the main text now follow by assuming all
unknowns are independent on time.

\section{Validation of the MT length distribution as measured in the simulations}\label{app:distrib}
We compare the MT length distribution as measured in the simulations to the analytical predictions following from Eqs.\ (\ref{eq:l_density}) for both nucleation scenarios in model M0. Figure \ref{fig:M0_distribution} shows the high level of agreement achieved. Error bars in these simulations are smaller than the plotting symbols. That the order parameter values are then also accurately reproduced is shown in the right panel of Fig.\ \ref{fig:M0_size}. 

\begin{figure}[h]
\centerline{\includegraphics[width=0.7\textwidth]{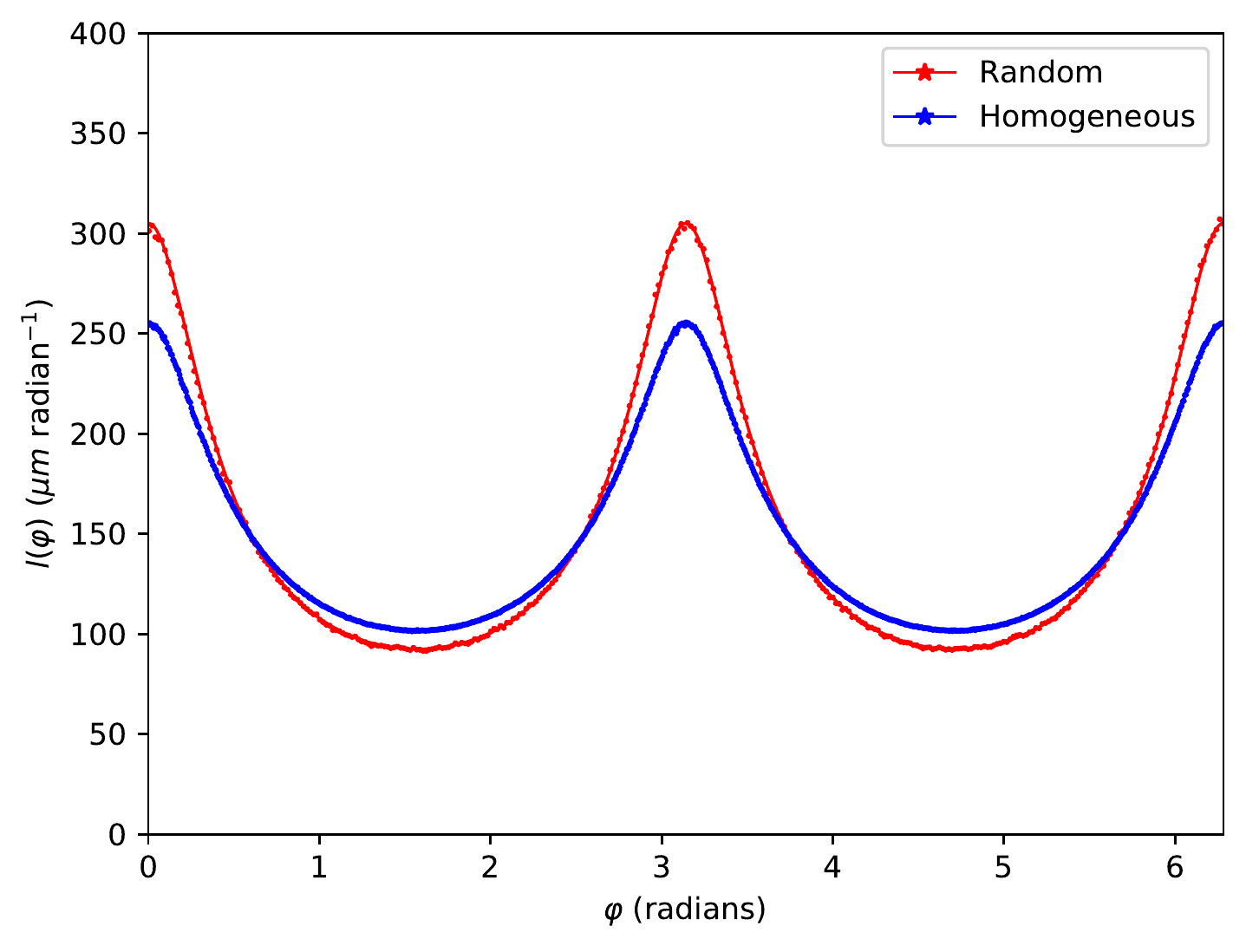}}
\caption{\textbf{Model M0}: Length distribution $l(\protect\varphi)$. Solid
lines: theoretical prediction, colored symbols: simulations, for the random
nucleation scenario (red symbols) and the the homogeneous scenario (blue
symbols). Parameters: long semi-axis $b=$\SI{4}{\micro\meter}, $r_{\mathrm{u}}=$\SI{0.01}{\per\second}, average taken over $20\cdot 10^6$ time steps.\label{fig:M0_distribution}}
\end{figure}

\section{PF density-dependent catastrophe rate in model MSP}
\label{app:r+_MSP}
We implement the influence of the PFs on the force-dependent catastrophe
mechanism of model MS through the unloaded catastrophe rate $r_{+}$ by
making it dependent on the PF density through imposing the constraint Eq.\ %
\ref{eq:tau_c_cb}. To that end we use the previously determined for the
compression modulus $k=$\SI{0.3}{\pico\newton\per\micro\metre} and unloaded growth speed $v_{+}=$\SI{0.018}{\micro\metre\per\second}, and take the unloaded catastrophe rate $r_{+}(c_{\mathrm{b}}=0)=$\SI{0.0078}{\per\second}, which matches the value of $r_{\mathrm{u}}=$\SI{0.01}{\per\second}, i.e\ $%
\langle\tau_{\mathrm{c}}\rangle(c_{\mathrm{b}}=0)=$\SI{100}{\second}, used in model MP. Using these values, we then solve Eq.\ \ref{eq:tau_c_cb} for a range of $c_{\mathrm{b}}$ values.  We show the results after translating the density to a number of PFs perlength-bin on the surface in the simulation for the case $c_* = 60$ in Fig.\ \ref{fig:r+_MSP}. 
\begin{figure}[h]
\centerline{\includegraphics[width=0.75\textwidth]{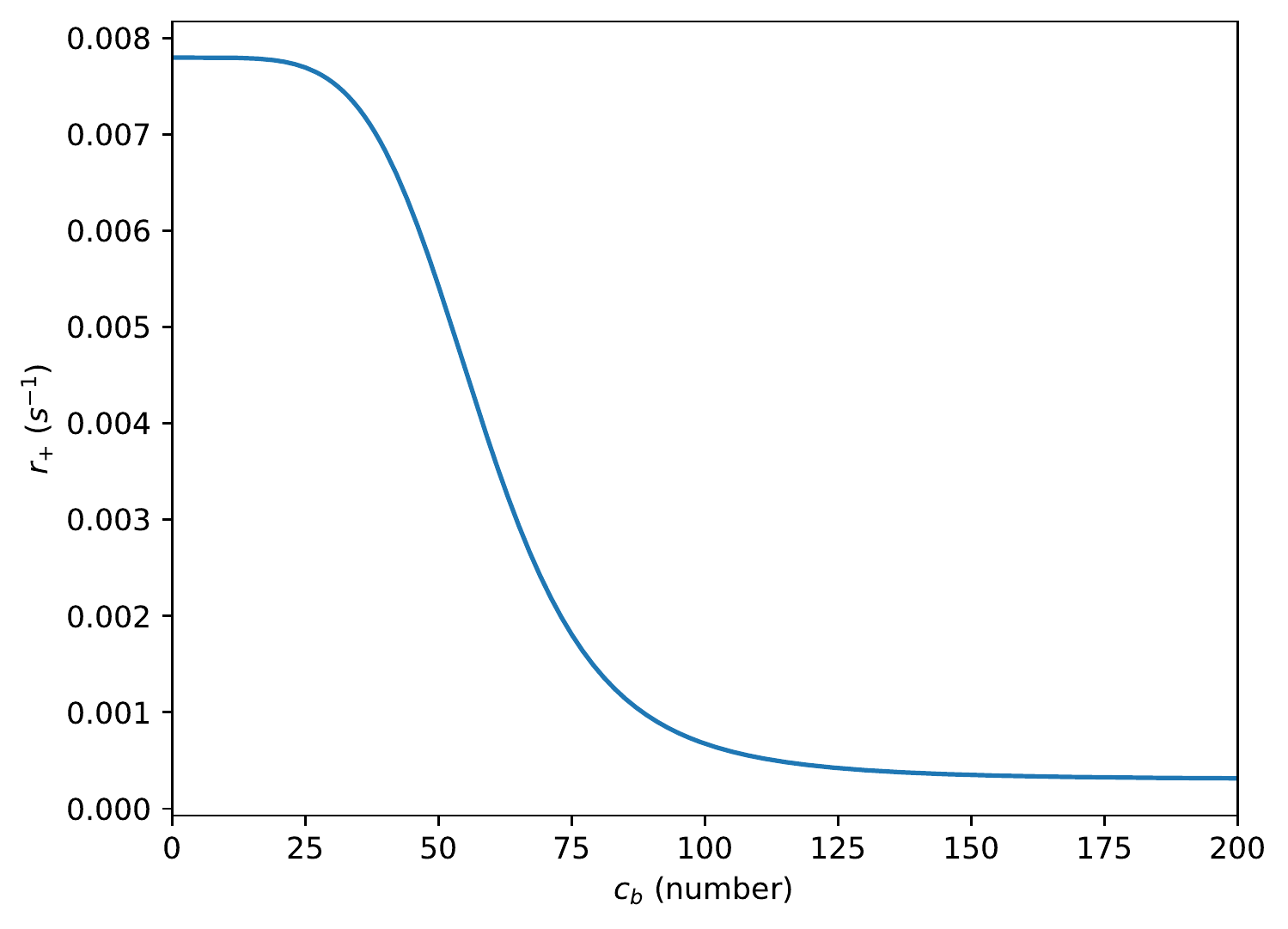}}
\caption{Force-free catastrophe rate $r_{+}$ (in \SI{}{\per\second}) as a function of
the number of polarity factors $c_{\mathrm{b}}$ per simulation bin at the cell
boundary. \label{fig:r+_MSP}}
\end{figure}

\section{Toy-model of the polarization mechanism \label{app:toy}}

\subsection{Assumptions}

We develop a simple and analytically tractable toy model to aid the analysis
of the results of the two models that involve the polarization mechanism,
Model MP (Section \ref{sec:MP}) and Model MSP (Section \ref{sec:MSP}). The
first simplifying assumption is to focus exclusively on the competition
between the shorter transverse axis, with length scale $a$ and the longer
longitudinal axis with length scale $b>a$. Instead of considering
isotropically nucleated MTs, we therefore consider a discrete direction
model where MTs are only nucleated in the directions $\varphi=0$ and $%
\varphi=\pi$, corresponding to the longitudinal direction, and $%
\varphi=\pi/2 $ and $\varphi=3\pi/2$, corresponding to the transverse
direction. The second assumption is that if the diffusion length of the PFs
in the membrane, which is given by $\lambda=\sqrt{D/k_u}$ is small compared
to a quarter of the circumference of the cell, we can neglect the diffusional
cross-talk between PFs delivered at different sites because the membrane is
closed. MTs in each of the discrete direction thus deliver their PFs to
their own unbounded membrane, from which they can subsequently unbind to
return to the cell interior. In case the sliding mechanism is also present,
we assume that all MTs rapidly slide to the poles, so that effectively we
only need to consider MTs nucleated in the two longitudinal directions.
Finally, instead of fixing the total number of MTs, we only specify their
rate of nucleation, which we take to be isotropic, i.e.\ corresponding to
the homogeneous nucleation scenario. Although not essential, this latter
assumption greatly simplifies the analysis. This toy model is illustrated in
Figure \ref{fig:toy_model}. 
\begin{figure}[tbp]
\centering
\includegraphics[width=\textwidth]{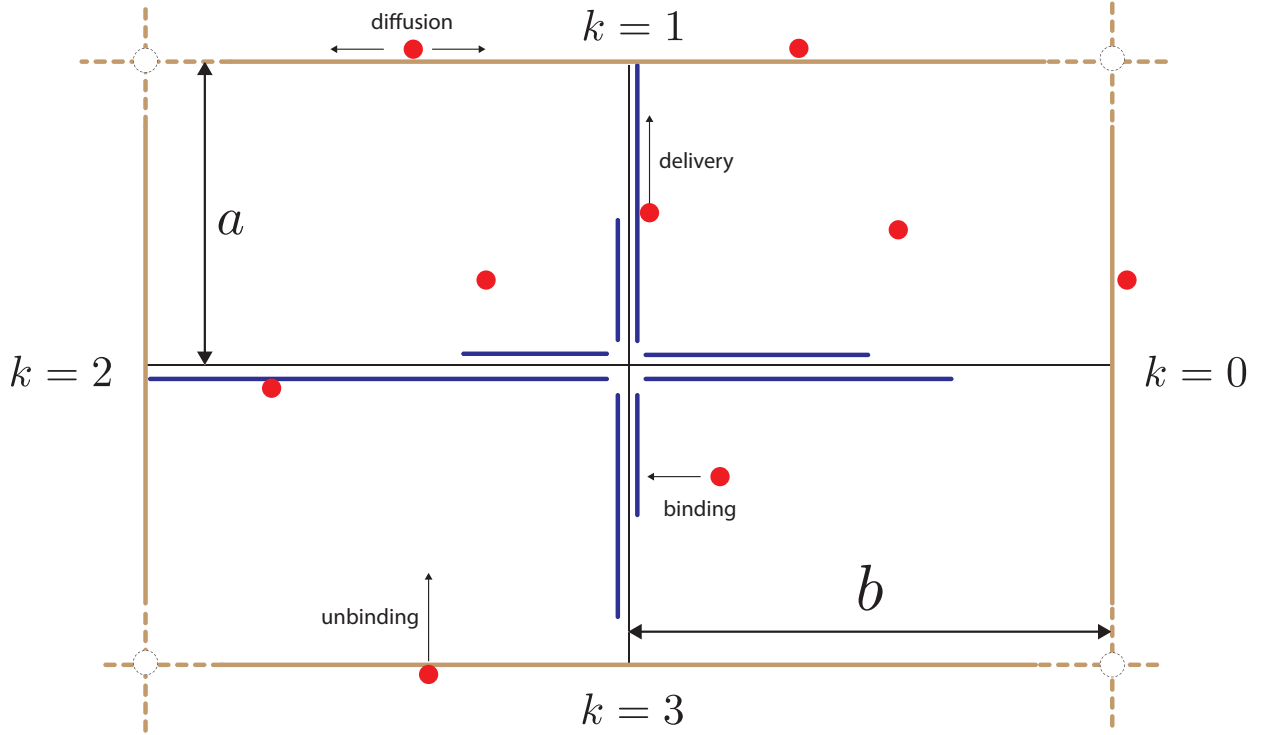}
\caption{Schematic of the toy model for the polarization mechanism. \label{fig:toy_model}}
\end{figure}

\subsection{General formulation}

In general our toy-model can have $N$ different discrete directions labelled
by an index $i=0,1,\ldots N-1$, corresponding to the spatial directions 
$\varphi =\frac{2\pi }{\mathrm{n}}i$. The distance from the central MTOC to the
boundary in the different directions is given by $d_{i}$. Assuming we are in
steady-state, the relevant variables are the MT length-densities $m_{i}^{\pm
}(l)$, the number of membrane-bound MTs $M_{i}^{\mathrm{b}}$ and the local
PF-densities $c_{i}^{\mathrm{b}}(s_{i})$, where $s_{i}$ is a signed distance
coordinate in the membrane connected to the direction $i$. The unbinding
rate of MTs is given by Eq.~(\ref{eq:ru_P}) evaluated in $c_{i}^{\mathrm{b}}(s_{i}=0)$%
. The nucleation rate of new MTs is given by $r_{\mathrm{n}}$ \emph{per direction}.
Denoting the value of the MT unbinding rate by $r_{u,i}$, which we note is a quantity that needs to be self-consistently determined, the solutions of the steady-state MT densities is simply given by (cf.\ Appendix \ref{app:model_M0} and \ref{sec:M0}) 
\end{subequations}
\begin{align}
m_{i}^{+}\left( l\right) & =\frac{r_{\mathrm{n}}}{v_{+}}e^{-l/\bar{l}} \\
m_{i}^{-}\left( l\right) & =\frac{r_{\mathrm{n}}}{v_{-}}e^{-l/\bar{l}} \\
M_{i}^{\mathrm{b}}& =\frac{v_{+}}{r_{u,i}}m_{i}^{+}\left( d_{i}\right) =\frac{r_{\mathrm{n}}}{%
r_{u,i}}e^{-d_{i}/\bar{l}}\equiv \frac{r_{\mathrm{n}}}{r_{u,i}}F_{i}.
\label{eq:toy_Mb}
\end{align}%
A key quantity in our model is the total length of MTs in
the direction $i$%
\begin{align}
L_{i}& =\int_{0}^{d_{i}}\mathrm{d}l\,l\left\{ m_{i}^{+}\left( l\right)
+m_{i}^{-}\left( l\right) \right\} +d_{i}M_{i}^{\mathrm{b}} \\
& =r_{\mathrm{n}}\bar{t}\left( \bar{l}\left( 1-F_{i}\right) -d_{i}F_{i}\right)
+d_{i}M_{i}^{\mathrm{b}},
\end{align}%
and its total $L=\sum_{i}L_{i}$.

Turning to the PF dynamics, we first define the total number of bound PFs
per direction 
\begin{equation}
C_{i}^{\mathrm{b}}=\int_{-\infty }^{\infty }ds_{i}\,c_{i}^{\mathrm{b}}(s_{i}),
\end{equation}%
and the total number of bound PFs $C^{\mathrm{b}}=\sum_{i}C_{i}^{\mathrm{b}}$. The density per unit length of MT-bound PFs in the cell the follows from Eq. (\ref{eq:cm}) 
\begin{equation}
c^{\mathrm{m}}=\frac{1}{L+L_{\frac{1}{2}}}(C-C^{\mathrm{b}}).
\end{equation}
The net flux of PFs reaching the membrane in the direction $i$ is thus%
\begin{equation}
K_{i}^{\mathrm{b}}=v_{\mathrm{m}}c^{\mathrm{m}}M_{i}^{\mathrm{b}}.
\end{equation}%
In steady state, the PF-density in this membrane satisfies 
\begin{equation}
D\frac{d^{2}}{ds^{2}}c_{i}^{\mathrm{b}}\left( s\right) -k_{\mathrm{u}}c_{i}^{\mathrm{b}}\left( s\right)
+K_{i}^{\mathrm{b}}\delta \left( s\right) =0,
\end{equation}%
with solution%
\begin{equation}
c_{i}^{\mathrm{b}}\left( s\right) =\frac{1}{2}\frac{K_{i}^{\mathrm{b}}}{\ell k_{\mathrm{u}}}e^{-|s|/\ell },
\end{equation}%
where the free diffusion length is given by $\ell =\sqrt{D/k_{\mathrm{u}}}$. The
total number of PFs in this membrane is then simply $%
C_{i}^{\mathrm{b}}=K_{i}^{\mathrm{b}}/k_{\mathrm{u}}$ and $c_{i}^{\mathrm{b}}\left( 0\right) =\frac{1}{2}C_{i}^{\mathrm{b}}/\ell$.

We now non-dimensionalize by choosing $r_{\mathrm{n}}^{-1}$ as unit of time and $%
\bar{l}$ as unit of length, introducing $\Lambda =L/\bar{l}$ , $\delta
_{i}=d_{i}/\bar{l}$, $\bar{\tau}=\bar{t}r_{\mathrm{n}}$. We also introduce $C_{\ast
}, $ the cross-over number of PFs through $c_{\ast }=\frac{1}{2}C_{\ast
}/\ell , $ and use this as a unit of measurement for PFs, introducing $%
\Gamma_{i}=$ $C_{i}^{\mathrm{b}}/C_{\ast }=c_{i}^{\mathrm{b}}\left( 0\right) /c_{\ast }$.
This allows us to write%
\begin{equation}
\frac{K_{i}^{\mathrm{b}}}{k_{\mathrm{u}}C_{\ast }}=\omega \frac{1}{\Lambda \left( \left\{
\Gamma_{j}\right\} \right) +\Lambda _{\ast }}\left( \Gamma -\sum \Gamma
_{j}\right) M_{i}^{\mathrm{b}}\left( \Gamma_{i}\right) =\Gamma_{i},
\label{eq:SC1pre}
\end{equation}%
where $\omega =v_{\mathrm{m}}/(k_{\mathrm{u}}\bar{l})$. We note that 
\begin{equation}
\Lambda _{j}\left( \Gamma_{j}\right) =\bar{\tau}\left( \left(
1-F_{i}\right) -\delta _{i}F_{i}\right) +\delta _{i}M_{i}^{\mathrm{b}}\left( \Gamma
_{j}\right),
\end{equation}%
so that%
\begin{equation}
\Lambda \left( \left\{ \Gamma_{j}\right\} \right) =\bar{\tau}\left( \sum_{i}%
\left( 1-F_{i}\right) -\delta _{i}F_{i}\right) +\sum_{i}\delta
_{i}M_{i}^{\mathrm{b}}\left( \Gamma_{i}\right).
\end{equation}%
Introducing $\Lambda _{0}=\bar{\tau}\left( \sum_{i}\left( 1-F_{i}\right)
-\delta _{i}F_{i}\right) +\Lambda _{\ast }$ we can rewrite (\ref{eq:SC1pre})
as%
\begin{equation}
\omega \frac{M_{i}^{\mathrm{b}}\left( \Gamma_{i}\right) }{\Lambda
_{0}+\sum_{j}\delta _{j}M_{j}^{\mathrm{b}}\left( \Gamma_{j}\right) }=\frac{\Gamma
_{i}}{\Gamma -\sum \Gamma_{j}}.
\end{equation}%
Multiplying by $\delta_{i}$ and summing allows us to solve for%
\begin{equation}
\sum_{j}\delta _{j}M_{j}^{\mathrm{b}}\left( \Gamma_{j}\right) =\frac{\Lambda
_{0}\sum_{j}\delta _{j}\Gamma_{j}}{\omega \Gamma -\sum_{j}\left( \omega
+\delta _{j}\right) \Gamma_{j}}.
\end{equation}%
This in turn allows us to solve for%
\begin{equation}
M_{i}^{\mathrm{b}}\left( \Gamma_{i}\right) =\Lambda _{0}\frac{\Gamma_{i}}{\omega
\Gamma -\sum_{j}\left( \omega +\delta _{j}\right) \Gamma_{j}}.
\label{eq:MbGam}
\end{equation}%
Next, in Eq. (\ref{eq:ru_P}) we introduce $\rho =r_{\mathrm{u}}\left( 0\right)
/r_{\mathrm{u}}\left( \infty \right) >1$ and write%
\begin{equation}
r_{\mathrm{u}}\left( c_{i}^{\mathrm{b}}\left( 0\right) \right) =r_{\mathrm{u}}\left( \infty \right) 
\frac{\rho +\Gamma_{i}^{p}}{1+\Gamma_{i}^{p}}\equiv r_{\mathrm{u}}\left( \infty
\right) R\left( \Gamma_{i}\right).
\end{equation}%
This allows us to compactly formulate the remaining boundary conditions Eq.\ (%
\ref{eq:toy_Mb})%
\begin{equation}
\frac{r_{\mathrm{n}}}{r_{\mathrm{u}}\left( \infty \right) }F_{i}=R\left( \Gamma_{i}\right)
M_{i}^{\mathrm{b}}\left( \Gamma_{i}\right).  \label{eq:SC2pre}
\end{equation}%
Using this to eliminate $M_{i}^{\mathrm{b}}\left( \Gamma_{i}\right) $ from (\ref%
{eq:MbGam}), then yields our final equations%
\begin{equation}
U_{i}=\frac{R\left( \Gamma_{i}\right) \Gamma_{i}}{\omega \Gamma
-\sum_{j}\left( \omega +\delta _{j}\right) \Gamma_{j}},  \label{eq:SC}
\end{equation}%
where the constants on the left--hand side are given by 
\begin{equation}
U_{i}=\frac{r_{\mathrm{n}}}{\Lambda _{0}r_{\mathrm{u}}(\infty )}F_{i}>0.
\end{equation}

For the further analysis it is useful to define the denominator of (\ref%
{eq:SC}) as a separate function $W\left( \left\{ \Gamma_{j}\right\} \right)
=\omega \Gamma -\sum_{j}\left( \omega +\delta _{j}\right) \Gamma_{j}$ and
note that is satisfies the bounds $0<$ $W\left( \left\{ \Gamma_{j}\right\}
\right) \leq \omega \Gamma $ and is monotonically decreasing in each of its
variables in the physical domain $\Gamma_{j}\geq 0.$ The function in the
numerator $R\left( \Gamma \right) \Gamma \geq 0$ has an inflexion point and
two local extrema whenever $\rho >\left( \left( p+1\right) /\left(
p-1\right) \right) ^{2}$ and diverges as $\Gamma \rightarrow \infty ,$ so
its inverse can be multivalued over a finite range.

\subsection{Models TOY-MP and TOY-MSP}
In the case intended to mimic Model MP, we have four discrete directions,
corresponding to the two longitudinal
orientations $\varphi =0$ and $\varphi =\pi $ and the two transverse
directions $\varphi =\pi /2$ and $\varphi =3\pi /2$. We define $\beta
=\delta _{0}=\delta _{2}=b/\bar{l}$ and $\alpha =\delta _{1}=\delta _{3}=a/%
\bar{l}$, $U_{\beta }=U_{0}=U_{2}$ and $U_{\alpha }=U_{1}=U_{3},$ noting
that $U_{\beta }<U_{\alpha }$ as $b>a,$ and $W\left( \Gamma_{0}+\Gamma
_{2},\Gamma_{1}+\Gamma_{3}\right) =\omega \Gamma -\left( \omega +\beta
\right) \left( \Gamma_{0}+\Gamma_{2}\right) -\left( \omega +\alpha \right)
\left( \Gamma_{1}+\Gamma 3\right) .$ We can then write the self-consistency
equations\ (\ref{eq:SC}) in this case as%
\begin{eqnarray}
W\left( \Gamma_{0}+\Gamma_{2},\Gamma_{1}+\Gamma_{3}\right) U_{\beta }
&=&R\left( \Gamma_{0}\right) \Gamma_{0}=R\left( \Gamma_{2}\right) \Gamma
_{2} \\
W\left( \Gamma_{0}+\Gamma_{2},\Gamma_{1}+\Gamma_{3}\right) U_{\alpha }
&=&R\left( \Gamma_{1}\right) \Gamma_{1}=R\left( \Gamma_{3}\right) \Gamma
_{3}.
\end{eqnarray}%
WLOG, we can also require $\Gamma_{0}\geq \Gamma_{2}$ and $\Gamma_{1}\geq
\Gamma_{3}$ as this simply divides out the multiplicity due to the trivial
interchange symmetries $\Gamma_{0}\leftrightarrow \Gamma_{2}$ and $\Gamma
_{1}\leftrightarrow \Gamma_{3}$. 

This allows the following systematic. algorithm to find all possible solutions. 
\begin{itemize}
    \item Choose a $w\in \left[ 0,\omega
\Gamma \right] .$ Solve%
\begin{align}
R\left( \Gamma_{0}\right) \Gamma_{0}& =R\left( \Gamma_{2}\right) \Gamma
_{2}=wU_{\beta } \\
R\left( \Gamma_{1}\right) \Gamma_{1}& =R\left( \Gamma_{3}\right) \Gamma
_{3}=wU_{\alpha }
\end{align}%
As $R\left( \Gamma \right) \Gamma $ is a universal function, which is either
monotonic, or has two local extrema, and values on $[0,\infty )$ these two
equations always have solutions, which can in principle be of the types (i) $%
\Gamma_{0}=\Gamma_{2}$ and $\Gamma_{1}=\Gamma_{3},$ the default biaxial
reference solution, (ii) $\Gamma_{0}=\Gamma_{2}$ and $\Gamma_{1}>\Gamma
_{3}$, which we call transversely polarized, (iii) $\Gamma_{0}>\Gamma_{2}$
and $\Gamma_{1}=\Gamma_{3},$ which we call longitudinally polarized and
finally (iv) $\Gamma_{0}>\Gamma_{2}$ and $\Gamma_{1}>\Gamma_{3},$ which
we would call doubly polarized. Note that in case of multiple solutions of $%
R\left( \Gamma \right) \Gamma =$constant, we discard the unstable middle
solution for which $R^{\prime }\left( \Gamma \right) \Gamma +R\left( \Gamma
\right) <0$. 
\item Next we check if $W\left( \Gamma_{1}+\Gamma_{3},\Gamma
_{2}+\Gamma_{3}\right) =w.$ If yes, a self-consistent solution is found, if
not, choose another $w.$ Since the solutions $\left( \Gamma_{1}\left(
w\right) ,\Gamma_{3}\left( w\right) ,\Gamma_{2}\left( w\right) ,\Gamma
_{4}\left( w\right) \right) $ are readily determined, this procedure boils
down to the one dimensional self-consistency problem%
\begin{equation}
w=W\left( \Gamma_{1}\left( w\right) +\Gamma_{3}\left( w\right) ,\Gamma
_{2}\left( w\right) +\Gamma_{3}\left( w\right) \right) 
\end{equation}%
on $w\in \left[ 0,\omega \Gamma \right]$, which is guaranteed to have one
solution (the default solution (i) above), but may have more. 
\end{itemize}

As order parameters we take the discrete analogs of the order parameters defined in Section \ref{sec:geom_dynamics}, i.e.
\begin{align}
    \mathbf{S}_{1,x} &= \frac{\sum_{i=0}^{3} L_{i}\cos{\frac{\pi}{2}i}}{\sum_{i=0}^{3} L_{i}}=\frac{L_0-L_2}{L_0+L_1+L_2+L_3} \\
    \mathbf{S}_{1,y} &= \frac{\sum_{i=0}^{3} L_{i}\sin{\frac{\pi}{2}i}}{\sum_{i=0}^{3} L_{i}}=\frac{L_1-L_3}{L_0+L_1+L_2+L_3} \\
    S_2 &=\frac{\sum_{i=0}^{3} L_{i}\cos{\pi i}}{\sum_{i=0}^{3} L_{i}}=\frac{(L_0+L_2)-(L_1+L_3)}{L_0+L_1+L_2+L_3}.
\end{align}

In the case intended to mimic Model MSP we have only two directions,
corresponding to the two longitudinal orientations $\varphi =0$ and $\varphi
=\pi $. In this case we have $W\left( \Gamma_{0}+\Gamma_{2}\right)
=\omega \Gamma -(\omega+\beta)\left( \Gamma_{0}+\Gamma_{2}\right) ,$ with $\Omega =\omega
\Gamma /\left( \omega +\beta \right) $ and $V=\left( \omega +\beta \right)
U_{\beta }$, and the self-consistency equations become%
\begin{equation}
W\left( \Gamma_{0}+\Gamma_{2}\right)U_{\beta}=R\left( \Gamma_{0}\right) \Gamma
_{0}=R\left( \Gamma_{2}\right) \Gamma_{2} . \label{eq:SC_TOY_MSP}
\end{equation}%
The only order parameter relevant to this case is $\mathbf{S}_{1,x}$, as $ \mathbf{S}_{1,y}=0$ and $S_2 =1$ by construction.

From symmetry it is clear that (\ref{eq:SC_TOY_MSP}) admits a biaxial
`reference' solution of the form $\Gamma ^{\left( 0\right) }=\Gamma
_{0}=\Gamma_{2}$ satisfying%
\begin{equation}
W\left( 2\Gamma ^{\left( 0\right) }\right) V=R\left( \Gamma ^{\left(
0\right) }\right) \Gamma ^{\left( 0\right)}.
\end{equation}%
Although we would like to study the solutions of this equation as a function
of $\Omega ,$ which through its linear dependence on $\Gamma $ is a proxy
for the total amount of PFs in the system, it is actually simpler to study
the inverse problem, and consider%
\begin{equation}
\Omega \left( \Gamma ^{\left( 0\right) }\right) =\frac{1}{V}\left\{ 2V\Gamma
^{\left( 0\right) }+R\left( \Gamma ^{\left( 0\right) }\right) \Gamma
^{\left( 0\right) }\right\}.
\end{equation}%
Taking the derivative with respect to $\Gamma ^{\left( 0\right) }$, here
denoted by a prime, we find%
\begin{equation}
\Omega ^{\prime }\left( \Gamma ^{\left( 0\right) }\right) =\frac{1}{V}%
\left\{ 2V+\left( R\left( \Gamma ^{\left( 0\right) }\right) \Gamma ^{\left(
0\right) }\right) ^{\prime }\right\}.
\end{equation}%
As $V>0$ this shows that $\Gamma ^{\left( 0\right) }\left( \Omega \right) $
can only be multivalued if $R\left( \Gamma ^{\left( 0\right) }\right) \Gamma
^{\left( 0\right) }$ is non-monotonic. 

\bibliographystyle{apsrev4-1}
\bibliography{cellgeometry}

\end{document}